\documentstyle[12pt,cite,epsfig]{article}
\def\lsim{\mathrel{\raise.3ex\hbox{$<$\kern-.75em\lower1ex\hbox{$\sim$}}}}
\def\gsim{\mathrel{\raise.3ex\hbox{$>$\kern-.75em\lower1ex\hbox{$\sim$}}}}
\def\beq{\begin{equation}}
\def\eeq{\end{equation}}

\newcommand{\be}{\begin{equation}}
\newcommand{\ee}{\end{equation}}
\begin{document}
\topmargin 0pt
\oddsidemargin=-0.4truecm
\evensidemargin=-0.4truecm
\renewcommand{\thefootnote}{\fnsymbol{footnote}}
\newpage
\setcounter{page}{1}
\begin{titlepage}     
\vspace*{-2.0cm}
\begin{flushright}
\vspace*{-0.2cm}
\end{flushright}
\begin{center}
{\Large \bf $(3+1)$-spectrum of neutrino masses: A chance for LSND?}
\vspace{0.5cm}

{O. L. G. Peres$^{1,2}$~\footnote{Permanent address:
Instituto de F\' {\i}sica Gleb Wataghin, UNICAMP, Campinas, Brazil.}
and  A. Yu. Smirnov$^{1,3}$\\
{\em (1) The Abdus Salam International Centre for Theoretical Physics,  
I-34100 Trieste, Italy }\\
{\em (2) Instituto de F\' {\i}sica Gleb Wataghin, 
Universidade Estadual de Campinas, UNICAMP
13083-970 Campinas SP, Brazil}\\
{\em (3) Institute for Nuclear Research of Russian Academy 
of Sciences, Moscow 117312, Russia}

}
\end{center}


\def\lsim{\mathrel{\raise.3ex\hbox{$<$\kern-.75em\lower1ex\hbox{$\sim$}}}}
\def\gsim{\mathrel{\raise.3ex\hbox{$>$\kern-.75em\lower1ex\hbox{$\sim$}}}}
\def\beq{\begin{equation}}
\def\eeq{\end{equation}}


\vglue 0.8truecm

\begin{abstract}  

If active to active neutrino transitions are dominant modes of the 
atmospheric ($\nu_{\mu} \rightarrow \nu_{\tau}$) and the solar 
neutrino oscillations ($\nu_{e}\rightarrow  \nu_{\mu}/\nu_{\tau}$), as is
indicated by recent data,  the favoured scheme which accommodates the LSND
result -- the so called $(2+2)$-scheme -- should be discarded. 
We introduce  the parameters $\eta_s^{atm}$ and  $\eta_s^{sun}$ 
which quantify an involvement of the sterile component in the solar and 
atmospheric neutrino oscillations. 
The $(2+2)$-scheme predicts $\eta_s^{atm} + \eta_s^{sun} = 1$ and 
the experimental  proof of  
deviation from this  equality will discriminate the scheme.  In this
connection the $(3+1)$-scheme is revisited in which the fourth
(predominantly sterile)
neutrino is isolated from a block of three flavour neutrinos  by the mass
gap $\Delta m^2_{LSND} \sim (0.4 - 10)$~eV$^2$. We find that in the
$(3+1)$-scheme
the LSND result can be reconciled with existing bounds 
on $\nu_e$- and $\nu_{\mu}$ - disappearance at $95-99 \%$ C.L.. 
The generic prediction of the scheme is the 
$\nu_e$- and $\nu_{\mu}$ - disappearance probabilities at the level of
present experimental bounds. 
The  possibility  
to strengthen the bound on $\nu_{\mu}$- disappearance 
in the KEK -- front detector  experiment is studied. We consider 
phenomenology of the 
$(3 + 1)$-scheme, in particular,  its
implications for the atmospheric neutrinos, neutrinoless 
double beta decay searches,
supernova neutrinos and primordial nucleosynthesis.

\end{abstract} 

\end{titlepage}
\renewcommand{\thefootnote}{\arabic{footnote}}
\setcounter{footnote}{0}
\newpage

\section{Introduction}
\label{sec:00a}

It is widely accepted that simultaneous explanation of the 
solar~\cite{solar,SK00}, atmospheric~\cite{SK00,SOUDAN00,macro00} and 
LSND results~\cite{lsnd-old,eitel,lsnd2000} in 
terms of neutrino conversion requires an existence of the  
sterile state~\cite{giunti,barger}. Less appreciated and
discussed fact is that explanation of the LSND result in the favored
$4\nu$-schemes implies that the solar or atmospheric neutrinos 
(or both) are converted to sterile
state. Indeed,  it is claimed~\cite{giunti,barger} that the only 
$4\nu$-schemes which can explain the LSND result and accommodate
oscillation solutions of the solar and atmospheric neutrino problems 
are the schemes of the (2 + 2) type. According to the (2 + 2) -
scheme, four neutrino mass eigenstates
form two nearly degenerate pairs separated by the gap 
$\Delta m^2=\Delta m^2_{LSND}\sim 1$~eV$^2$. Splittings between
masses within pairs are much smaller: they are determined by $\Delta
m^2_{atm}\approx 3 \cdot 10^{-3}$ ~eV$^2$ in the atmospheric neutrino
pair
and $\Delta m^2_{\odot} \lsim 10^{-4}$~eV$^2$ in the solar neutrino
pair. 

The key feature of the (2 + 2)-scheme which allows one to get a large
enough
probability of the $\bar{\nu}_{\mu} \rightarrow \bar{\nu}_e$ transition
and to 
satisfy existing bounds  on mixing parameters is that 
$\nu_e$ and  $\nu_{\mu}$ are present in different
pairs: $\nu_e$ is mainly distributed in the pair
with $\Delta m^2_{\odot}$ - splitting, 
whereas $\nu_{\mu}$ is mainly distributed 
in the pair with $\Delta m^2_{atm}$ - splitting. Indeed, the depth  of the
$\bar{\nu}_{\mu} \leftrightarrow  \bar{\nu}_e$  
oscillations driven by $\Delta m^2_{LSND}$ equals  
\begin{equation}
\sin^2 2 \theta_{LSND} =
4\, (\sum_{j=3,4} U_{\mu j} U_{e j})^2 = 
4\, (\sum_{j=1, 2} U_{\mu j} U_{e j})^2 
\sim \epsilon^2,
\label{eq:1a}
\end{equation}
where summation is over the mass eigenstates in the heavy 
(or light) degenerate pair. If we introduce small parameter 
$\epsilon$ which describes admixture
of $\nu_e$  (or $\nu_{\mu}$) in the pair where it is not a dominant
component, then taking into account that  in a given pair either
$U_{\mu j}\approx {\mbox O}(1)$ or 
$U_{ej} \approx {\mbox O} (1)$, we get that the effective mixing is of
the order $\epsilon^2$.

The sterile neutrino can be distributed in two pairs in different
ways, and  there are two extreme versions of the  $(2+2)$-scheme.
1) The sterile neutrino is mainly in the solar pair, so that  the
$\nu_e\rightarrow \nu_s$ conversion is responsible for the solution of
the solar neutrino problem, whereas atmospheric neutrino problem  is
solved by $\nu_{\mu}\leftrightarrow \nu_{\tau}$ oscillations.
2) The sterile neutrino is in the atmospheric pair, then 
$\nu_{\mu}$ oscillates into $\nu_{s}$ and  the solar
neutrino problem is solved by the $\nu_{e}\leftrightarrow \nu_{\tau}$
conversion. The intermediate situations are possible 
when both the solar and atmospheric neutrinos partly  oscillate 
to  the sterile component. 
The exceptional case is when $\nu_{s}$ is distributed
equally in the solar and atmospheric neutrino pairs. In all other
cases the sterile neutrino contributes either to  solar or to 
atmospheric neutrinos more than one half. 

\begin{figure}
\begin{center}
\parbox[c]{4.5in}
{\mbox{\qquad\epsfig{file=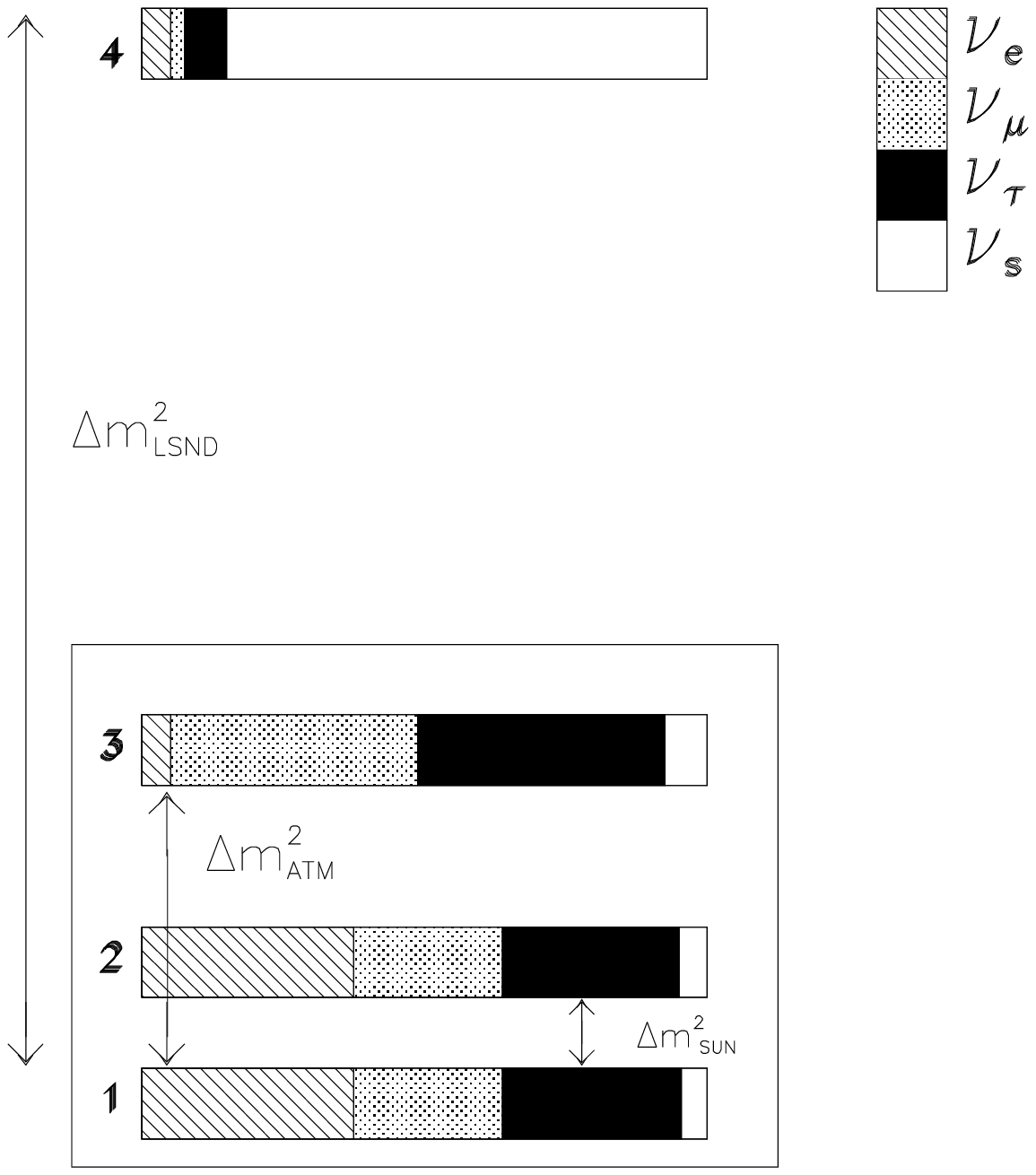,width=1.2\linewidth,height=1.2\linewidth}}}
\end{center}
\caption{
The neutrino mass and flavor spectrum in the $(3 +1)$ scheme 
with normal mass hierarchy. The boxes correspond to the mass eigenstates. 
The sizes of different regions in the boxes determine flavors of the
eigenstates: $|U_{\alpha i}|^2$ , $\alpha = e, \mu, \tau, s$, 
$i = 1, 2, 3, 4$. 
}
\label{xfig1}
\end{figure}

In the alternative $(3+1)$-scheme  ( see Fig.~\ref{xfig1}), three mass
eigenstates
form ``flavour block'' which has predominantly active flavour with
small admixture of the
sterile state and small mass splitting. The fourth mass eigenstate is 
separated from the flavour block by the LSND mass gap $\Delta
m^2_{LSND}\approx (0.4 - 10)$~eV$^2$; it consists mainly of the sterile
neutrino with small admixtures of active neutrinos: 
$|U_{s 4}|^2 \approx 1$,   $|U_{\alpha 4}|^2<<1$,
$\alpha=e,\mu$ and $\tau$. 
(In general, there is  mixing of $\nu_s$ and $\nu_{\tau}$ in 
$\nu_4$  and in the flavor block~\cite{giunti-31}.) 
In the $(3+1)$-scheme 
the atmospheric and solar neutrino data are explained by the
$\nu_{\mu} \rightarrow \nu_{\tau}$ oscillations and $\nu_{e} \rightarrow
\nu_{\mu}/\nu_{\tau}$ conversion respectively.

It is claimed that the $(3+1)$-mass scheme  can not reproduce 
large enough probability of the $\bar{\nu}_{\mu} \rightarrow \bar{\nu}_{e}$
oscillations 
to explain the LSND result~\cite{giunti,barger}. Indeed, the depth
of $\bar{\nu}_{\mu} \rightarrow \bar{\nu}_{e}$ oscillations driven by the 
$\Delta m^2_{LSND}$ equals
\begin{equation}
\sin^2  2 \theta_{LSND} = 
4\, U_{\mu 4}^2 U_{e 4}^2 \sim \epsilon^4 ,
\label{eq:1b}
\end{equation}
where $U_{\mu 4}$ and  $U_{e4}$, are the admixtures  of the 
$\nu_e$ and $\nu_{\mu}$ in the 4th mass eigenstate. These admixtures
are small ($\sim \epsilon$)  being restricted  by the accelerator
and reactor experiments. As a consequence, 
the depth of oscillations is of the fourth  order in the  small
parameter. 
Notice, however, that early 
analysis of data (with small statistics) showed that the
bounds from accelerator and reactors experiments 
and the positive LSND result 
can be reconciled  in the context of one level dominance 
scheme   at $99 \% $ C.L.~\cite{lisi-old}.

There are three recent results which may change eventually the
situation in favor of the $(3+1)$-scheme:

\begin{enumerate} 

\item The data on atmospheric neutrinos (specifically on
zenith angle distribution of the upward-going muons,
on the  partially contained multi-ring events,  with  $E_{\nu}>
5$~GeV, 
and on the enriched neutral current event sample)  disfavour 
oscillations to the sterile state~\cite{SK-sterile}. 
The $\nu_{\mu} \leftrightarrow \nu_s$ oscillations can be
accepted by the data at $3\sigma$ level. Oscillations of  active
neutrinos, $\nu_{\mu} \rightarrow \nu_{\tau}$, give better fit
of the experimental results.

\item Conversion of solar $\nu_{e}$ neutrinos  to active
neutrinos gives 
better global fit of experimental results than conversion to sterile
neutrinos~\cite{SK00}.

If active neutrino channels dominate  both in the solar and in atmospheric
neutrino oscillations, the $(2+2)$ -mass scheme should be discarded and the
oscillation 
interpretation of the LSND result becomes problematic. In this
connection, we have reconsidered the possibility to explain LSND
results in the $(3+1)$ -mass scheme~\cite{alei1}.

\item Latest  analysis of the LSND data~\cite{lsnd2000} 
shows in further shift  of the
allowed region of oscillation parameters to smaller values of mixing
angle. This leads to a better agreement
between the bounds obtained in  
CDHS~\cite{cdhs}, CCFR~\cite{ccfr85} and Bugey~\cite{bugey} 
experiments and the LSND result in the context
of $(3+1)$-mass scheme~\cite{giunti-31,barger-31,kayser-31}.

\end{enumerate}

In this paper we 
elaborate on the phenomenology of the $(3+1)$-scheme 
which can accommodate the oscillation 
interpretation of the LSND result.   
In sect.~2 we introduce parameters $\eta_s^{sun}$ and 
$\eta_s^{atm}$ which quantify participation of the sterile neutrino in the 
solar and atmospheric neutrino oscillations.
We show how  these parameters can be used to disentangle 
the (2 + 2)- and (3 + 1)-schemes.  
In sect.~3 we   analyze  the  $\bar{\nu}_{\mu} \rightarrow \bar{\nu}_e$ 
oscillation probability 
in the $(3+1)$-scheme. We estimate the confidence level for 
the bound on $\sin^2 2\theta_{e \mu}$ from $\nu_{\mu}$-and 
$\nu_{e}$-disappearance experiments. In sect.~4, we 
study a possibility to improve the 
bounds on   $\nu_{\mu}$-disappearance, and consequently, on
$\sin^2 2\theta_{e\mu}$  
with the KEK - front detectors experiments.
In sect.~5, we describe 
phenomenological implications of the $(3+1)$-scheme 
for atmospheric neutrinos, neutrinoless double beta decay searches, 
primordial nucleosynthesis and  
supernova neutrinos. In sect.~6 we 
summarize our results and discuss perspectives to identify 
$(3+1)$-mass scheme. In Appendix, we evaluate how strongly the
LSND oscillation probability can be enhanced in the schemes  with
more than  one sterile neutrino.

\section{The (2 +2)-scheme  versus (3 + 1)-scheme}

As we have pointed out in the introduction, the present data  disfavor pure 
$\nu_{\mu} \leftrightarrow \nu_s$ oscillations of the atmospheric
neutrinos and moreover 
pure $\nu_{e} \rightarrow \nu_s$ conversion is not the best solution of
the solar neutrino problem. Let us consider a general case when 
$\nu_s$ is involved  both in the solar and in the 
atmospheric neutrino oscillations. 

\subsection{Excluding the (2 +2)-scheme ?}

For definiteness we will consider the (2 + 2)-scheme with 
$\nu_1$ and $\nu_2$ being 
the ``solar" pair of the eigenstates with 
$ \Delta m^2_{\odot}\equiv m_2^2-m_1^2$ mass splitting and 
$\nu_3$ and $\nu_4$ -- the atmospheric pairs 
(with $ \Delta m^2_{atm}\equiv m_4^2-m_3^2$ mass splitting 
($m_1 < m_2 <  m_3 <  m_4$). 
To a good approximation (as far as solar and atmospheric neutrinos are
concerned) we can neglect small admixtures of 
$\nu_{e}$ and $\nu_{\mu}$ implied by the LSND result. So, the  
$\nu_{e}$-flavor is present in the solar pair: 
\begin{equation}
|U_{e1}|^2 + |U_{e2}|^2 \approx 1~, 
\label{eq:solarp}
\end{equation}
and the $\nu_{\mu}$ flavor is in the atmospheric pair: 
\begin{equation}
|U_{\mu 3}|^2 + |U_{\mu 4}|^2 \approx 1~. 
\label{eq:atmp}
\end{equation}

Under conditions (\ref{eq:solarp}) and (\ref{eq:atmp}) 
the effect of the sterile neutrino both in solar and in 
atmospheric neutrinos is described by a unique parameter~\cite{4solar-nu}. 
In what follows, we will introduce this parameter in a more transparent way.  
Notice that since both solar and atmospheric neutrino oscillations 
are reduced to 2$\nu$ cases, there is no CP-violation effect and the
absolute values of mixings are relevant only. 

In general, $\nu_{\tau}$ and $\nu_s$ are mixed both in the 
solar and atmospheric neutrino pairs. 
Using the
unitarity of the mixing matrix and the equality (\ref{eq:solarp}) 
it is easy to show  that in
the solar pair ($\nu_1, \nu_2$) the electron neutrino mixes with the 
combination: 
\begin{equation}
\tilde{\nu} = \cos \alpha ~\nu_s + \sin \alpha ~\nu_{\tau}, 
\label{tilde}  
\end{equation}
(where $\alpha$ is the arbitrary mixing angle) so that 
\begin{equation}
\nu_1 =  \cos \theta_{\odot}~ \nu_e -  \sin \theta_{\odot}
~\tilde{\nu},~~~
\nu_2 =  \cos \theta_{\odot} ~\tilde{\nu} + \sin \theta_{\odot}~ \nu_e~.   
\label{e-tilde}
\end{equation}
Here $\theta_{\odot}$ is the angle which  appears in the $2\nu$-analysis
of the solar neutrino data. 
Then as follows from the unitarity and the condition 
(\ref{eq:atmp}), the combination of $\nu_s$ and $\nu_{\tau}$ 
orthogonal to that in Eq. (\ref{tilde}): 
\begin{equation} 
\nu' = \cos \alpha ~\nu_{\tau} - \sin \alpha ~\nu_s,  
\label{prime}
\end{equation}
mixes with $\nu_{\mu}$ in the atmospheric neutrino pair 
($\nu_3, \nu_4$): 
\begin{equation}
\nu_3 =  \cos \theta_{atm} \nu_{\mu} -  \sin \theta_{atm}\nu' ,~~~
\nu_4 =  \cos \theta_{atm} \nu' + \sin \theta_{atm} \nu_{\mu},
\label{mu-prime}
\end{equation}  
where $\theta_{atm} \sim 45^0$ is the mixing angle responsible for
oscillations of the atmospheric neutrinos. 

Using Eqs. (\ref{tilde}, \ref{e-tilde}) we find 
\begin{equation}
\eta_s^{sun} \equiv |U_{s1}|^{2} + |U_{s2}|^{2} =  \cos^2 \alpha~, 
\label{etasun}
\end{equation}
that is,  $\cos^2 \alpha$  or $\eta_s^{sun}$ give  the total presence 
of the sterile 
neutrino in the solar pair  and describe the
effects of the sterile
component in the solar neutrinos. 
Similarly, from the Eq. (\ref{prime}, \ref{mu-prime}) we get 
\begin{equation}
\eta_s^{atm} \equiv |U_{s3}|^{2} + |U_{s4}|^{2} =  \sin^2 \alpha~, 
\label{etaatm}
\end{equation}
so, $\sin^2 \alpha$ or $\eta_s^{atm}$ 
give  a total presence of the sterile component 
in the atmospheric pair. 

Clearly, in the (2 + 2)-scheme: 
\begin{equation}
\eta_s \equiv \eta_s^{sun} + \eta_s^{atm} = 1. 
\label{eq:sum}
\end{equation}
The schemes  with pure $\nu_{e} \rightarrow \nu_s$ conversion of the solar
neutrinos and $\nu_{\mu} \leftrightarrow  \nu_{\tau}$ oscillations of the 
atmospheric neutrinos correspond to 
$\cos^2 \alpha = 1$. The opposite situation: the solar $\nu_{e} -
\nu_{\tau}$ and the atmospheric 
$\nu_{\mu} - \nu_s$ transitions corresponds to $\cos^2 \alpha = 0$. 

Notice that in notations of Ref. \cite{4solar-nu} 
$\cos^2 \alpha \equiv c_{23}^2 \cdot  c_{24}^2$.  

The equality (\ref{eq:sum}) is a generic property of the (2 + 2)-scheme
and it can be checked experimentally. 
One should measure or restrict $\eta_s^{sun}$ 
from the solar neutrino data,  
and independently,  $\eta_s^{atm} $ from the atmospheric neutrino data.  
According to  
Eqs. (\ref{etasun}) and  (\ref{tilde}),  
$\sqrt{\eta_s^{sun}}$ is the admixture of the 
$\nu_s$ in the state 
\begin{equation}
\tilde{\nu} = \sqrt{\eta_s^{sun}}~\nu_s + \sqrt{1 - \eta_s^{sun}}~
\nu_{\tau}
\label{tldeeta}
\end{equation}
to which the solar $\nu_{e}$ convert. 
Thus, $\eta_s^{sun}$ should be determined from the fit of the solar neutrino
data in terms of the 2$\nu$ mixing of $\nu_{e}$ and $\tilde{\nu}$. 
Similarly (see Eqs. (\ref{prime}), (\ref{etaatm})),  
$\sqrt{\eta_s^{atm}} $ is the admixture of the  
$\nu_s$ in the state
\begin{equation}
\nu' = \sqrt{\eta_s^{atm}}~\nu_{\tau} - \sqrt{1 - \eta_s^{atm}}~\nu_s
\label{primeeta}
\end{equation} 
to which the atmospheric neutrinos oscillate. Thus, $\eta_s^{atm}$  
can be determined from the 2$\nu$-analysis of the atmospheric neutrino data 
in terms of $\nu_{\mu}- \nu'$ mixing.
  
If it will be proven that 
\begin{equation}
\eta_s^{sun} + \eta_s^{atm} < 1
\label{ineqsum}
\end{equation}
(or larger than 1), the (2 + 2)-scheme should be discarded. Let us
summarize the present situation. The global analysis of the  
solar neutrino data~\cite{concha-n2000} shows that for large mixing
angle solutions (LMA): 
\begin{equation}
\eta_s^{sun} < 0.44  ~~~(90 \% ~{\rm C.L.})~, ~~~~    
\eta_s^{sun} < 0.72 ~~~(99 \% ~{\rm C.L.})~;
\label{boundl}
\end{equation}
for LOW solution: 
\begin{equation}
\eta_s^{sun} < 0.30  ~~~(90 \% ~{\rm C.L.})~, ~~~~    
\eta_s^{sun} < 0.77 ~~~(99 \% ~{\rm C.L.})~;
\label{boundl1}
\end{equation}
and for small mixing angle solution (SMA):
\begin{equation}
\eta_s^{sun} < 0.90 ~~~(90 \% {\rm C.L.})~,
\label{boundsun} 
\end{equation}
and no bound appears at the $99 \% {\rm C.L.}$.  
The bound from the SK results on atmospheric neutrino data  
is~\cite{bari-4nu}: 
\begin{equation}
\eta_s^{atm} < 0.67 ~~~(90 \% {\rm C.L.})~.
\label{boundatm}   
\end{equation}
One can get similar bound  from the fit of the zenith angle distribution
of events detected by MACRO~\cite{macro00now}: $\eta_s^{atm}<0.7$ at
90 \% {\rm C.L.}. 

Thus, taking the  LMA solution, we get 
\begin{equation}
\eta_s \equiv \eta_s^{sun} +  \eta_s^{atm} <  1.11 ~~~~~ 
(90 \% {\rm C.L.}),
\end{equation}
and for the SMA solution: $\eta_s < 1.57$  and for the LOW solution:
$\eta_s < 0.97$. That is, at the moment  the  (2 + 2)-scheme is well
acceptable. 
However the forthcoming solar and atmospheric neutrino experiments can
significantly strengthen this bound.

Let us consider dependences of various observable on 
$\eta_s^{sun}$ and $\eta_s^{atm}$. 

The solar neutrinos undergo $\nu_{e} \rightarrow \tilde{\nu}$ conversion. 
Difference of the $\nu_{e}$ and $\tilde{\nu}$ potentials in matter equals
\begin{equation}
V = \sqrt{2} G_F \left( n_e - \eta_s^{sun} \frac{n_n}{2}\right)~, 
\label{poten1}
\end{equation}
where $G_F$ is the Fermi coupling constant,  $n_e$ and $n_n$
are the concentrations of electrons and neutrons correspondingly. 
With the increase of $\eta_s^{sun}$ the potential decreases, this leads 
to a shift of the adiabatic edge of the suppression pit to smaller 
$\Delta m^2/E_{\nu}$ and to modification (weakening) of the Earth matter
effect~\cite{act-ster}. 

One expects an intermediate situation between pure active and pure sterile
cases.

Presence of the sterile component in the solar neutrino flux  modifies
also interactions of neutrinos in detectors. 
The reduced rate of the 
neutral current events [NC]  defined as the    
ratio of events with and without oscillations,  
$N_{NC} (osc)/N_{NC} (SSM)$,  decreases with increase of $\eta_s^{sun}$: 
\begin{equation}
{\rm [NC]} = \eta_s^{sun} \bar{P} + (1 - \eta_s^{sun}), 
\label{ncrate}
\end{equation}
here $\bar{P}$ is the effective ( averaged) survival probability.
Two remarks are in order. 

1). The probability $\bar{P} = \bar{P}(\eta_s^{sun})$ should be
calculated
with the effective potential (\ref{poten1}), although its  dependence on 
$\eta_s^{sun}$  is rather weak. 

2). The probability should be calculated for each specific variable 
separately (taking into account cross sections, 
energy thresholds etc.) so that 
$\bar{P}$ in Eq. (\ref{ncrate}) may differ from $\bar{P}$ which will
appear in the following formulas.

The double ratio - the ratio of the reduced rates of the neutral 
current  events  [NC] and
the charged current events ([CC] $\equiv N_{CC} (osc)/N_{CC}(SSM)$) 
changes as: 
\begin{equation}
\frac{{\rm [NC]}}{{\rm [CC]}} = \eta_s^{sun} + 
\frac{1 - \eta_s^{sun}}{\bar{P}}~. 
\label{eq:a}
\end{equation}
With increase of $\eta_s^{sun}$ the double ratio decreases from 
$1/\bar{P}$ to 1. The rate [NC] and the ratio 
[NC]/[CC] will be measured in the  SNO experiment~\cite{sno}.

The reduced charged current event rate [CC] 
($\equiv N_{CC}(osc)/ N_{CC}(SSM)$) can be written in terms of the
suppression factor for the $\nu-e$ event rate, 
$R_{\nu e} \equiv N_{\nu e}(osc)/ N_{\nu e}(SSM)$,  as 
\begin{equation}
{\rm [CC]} =  \frac{R_{\nu e}}
{1 -  r  (1 - \eta_s^{sun})( 1 - 1/\bar{P}) }~.
\label{ccrate}
\end{equation}
Here $r \equiv \sigma(\nu_{\mu} e)/ \sigma(\nu_{e} e)$ is the ratio 
of the cross-sections of the muon and the electron neutrino scattering on
the electron. The rate $R_{\nu e}$ is   
measured 
with high precision at Super-Kamiokande and it  will be also measured at
SNO.  
According to (\ref{ccrate}), with increase of  involvement of the
sterile neutrino the rate increases from 
$R_{\nu e}/(1 -  r ( 1 - 1/\bar{P}))$ for the pure active case 
to $R_{\nu e}$ for the pure sterile
neutrino case. 

The presence of the sterile  component  in the atmospheric 
neutrinos can be established by studies of the neutral current 
interaction rates. 
The suppression factor for the $\pi^0$-event rate  in the pure
sterile case
is expected to be about $\xi = 0.7 - 0.8$ ~\cite{vissani}. In general 
case the ratio of $\pi^0$ to $e$-like event rates is suppressed as 
\begin{equation}
\frac{N(\pi^0)}{N_{e}} = \frac{N^0(\pi^0)}{N^0_{e}} 
(1 - \eta_s^{atm} + \xi \eta_s^{atm})~,  
\label{pizero}  
\end{equation}
where $N^0(\pi^0)$ and $N^0_{e}$ are the numbers of events with 
and without oscillations. 
 
Appearance of the $\nu_{\tau}$ in oscillations of $\nu_{\mu}$ is 
suppressed by the factor $\cos^2 \alpha \equiv 1 - \eta_s^{atm}$. 
This  can be tested in the long base-line experiments.  
One can compare the mixing parameter $\sin^2 2\theta_{\mu \mu}$ 
extracted from studies of the $\nu_{\mu}$ -  disappearance and 
the parameter $\sin^2 2\theta_{\mu \tau}$ found from the 
$\nu_{\tau}$  - appearance experiments. In the (2 + 2)-scheme one expects 
\begin{equation}
\frac{\sin^2 2\theta_{\mu \tau}}{\sin^2 2\theta_{\mu \mu}} 
= 1 - \eta_s^{atm}. 
\label{apdisap}
\end{equation}

Also the zenith angle distribution 
of the of multi-ring events from the so called NC enriched 
sample~\cite{SK-sterile} is sensitive to $\eta_s^{atm}$. 

The Earth matter effect on atmospheric $\nu_{\mu}$ neutrino oscillations 
 depends on $\eta_s^{atm}$. The matter potential for 
$\nu_{\mu} - \nu'$ system equals: 
\begin{equation}
V = \sqrt{2} G_F \frac{n_n}{2} \eta_s^{atm}. 
\label{potst}
\end{equation}
The potential increases with $\eta_s^{atm}$, and one expects an 
intermediate situation between active and sterile 
neutrino cases~\cite{4atm-nu,bari-4nu}. 
For high energy upward going muons the suppression of the oscillation
effect increases with $\eta_s^{atm}$. The effect is more complicated for 
neutrinos crossing the core of the earth where the parametric enhancement
of oscillations may take place.\\ 


\subsection{The $(3 +1)$-scheme and mixing of  sterile neutrino}

In the $(3 +1)$-scheme  with 4th (isolated) mass eigenstate 
being predominantly sterile one, total involvement 
of the sterile neutrino  in the solar and atmospheric neutrino oscillations
can be very small. If the $\nu_{\tau}$ component in the $\nu_4$ is 
absent, we have from unitarity condition 
\begin{equation}
\sum_{i = 1}^{3} U_{si}^2 =  |U_{e 4}|^2 + |U_{\mu 4}|^2 < (3 - 10) \cdot
10^{-2}
\label{etasst}
\end{equation}
where the inequality corresponds to the CDHS and Bugey bounds. 

The situation is different if $\nu_{\tau}$ mixes with $\nu_s$ 
in the 4th eigenstate~\cite{giunti-31}. 
Now $\nu_4$ consists mainly of the combination 
\begin{equation}
\tilde{\nu} = \cos \beta~ \nu_s  +  
\sin \beta~ \nu_{\tau}
\label{tildestau}
\end{equation}
with small admixtures 
of $\nu_{e}$  and $\nu_{\mu}$ implied by the LSND experiment 
($\nu_4 \approx \tilde{\nu}$). 
The mixing in  the flavor block can be obtained from the 
one in the original (3 + 1)-scheme (without $\nu_s - \nu_\tau$  mixing)    
by substituting  $\nu_{\tau} \rightarrow \nu'$, where 
\begin{equation}
\nu' = \cos \beta~\nu_{\tau}  -  \sin \beta~\nu_s. 
\label{primestau}
\end{equation}
is the orthogonal combination to that in Eq. (\ref{tildestau}).  

An immediate consequence of the $\nu_{\tau}-\nu_s$ mixing is 
an appearance of $\nu_{\tau} \leftrightarrow \nu_s$ oscillations driven by 
$\Delta m_{LSND}^2$~\cite{giunti-31}. The depth of these oscillations equals 
$\sin^2 2\beta$. 

Due to the presence of $\nu_{\mu}$ in  $\nu_4$ described by $U_{\mu 4}$ 
the scheme leads also to 
$\nu_{\mu} \leftrightarrow \nu_{\tau}$ oscillations with $\Delta m^2 = 
\Delta m_{LSND}^2$  and the depth 
\begin{equation}
\sin^2 2\theta_{\mu \tau} \simeq  4 |U_{\mu 4}|^2 \sin^2 \beta = 
\sin^2 2\theta_{\mu \mu} \cdot \sin^2 \beta~, 
\label{depthmt}
\end{equation}
where $\sin^2 2\theta_{\mu \mu}$ describes the disappearance of
$\nu_{\mu}$. Eq. (\ref{depthmt}) allows  to get  the upper bound on
$\sin^2 \beta$.  Indeed, using $\sin^2 2\theta_{\mu \tau}(C/N)$ -- 
the upper bound on $\sin^2 2\theta_{\mu \tau}$ 
from the CHORUS~\cite{chorus} and NOMAD~\cite{nomad} experiments and  taking 
$\sin^2 2\theta_{\mu \mu}$  
at the  upper edge allowed by the CDHS experiment 
(as is implied by the LSND result) we get:   
\begin{equation}
\sin^2 \beta \leq \frac{\sin^2 2\theta_{\mu \tau}(C/N)}
{\sin^2 2\theta_{\mu \mu}(CDHS)}~. 
\label{betabound} 
\end{equation}
From  this formula we find that for $\Delta m^2 < 2$ ~eV$^2$ no 
bound appears; for $\Delta m^2  = 4,~6, ~ 8$ ~eV$^2$ we get 
$\sin^2 \beta < 0.32,~ 0.13, ~ 0.06$ correspondingly. \\

In the presence of $\nu_{\tau} - \nu_s$ mixing the solar 
$\nu_{e}$ will convert into the combination

\begin{equation}
\nu_{\mu} \cos \theta_{atm} + \nu_{\tau} \sin \theta_{atm}
\cos\beta - \nu_s \sin \theta_{atm} \sin\beta.
\end{equation} 
If SMA is the solution for solar neutrino problem then the above state 
is roughly $~~(\approx \nu_2)$. 

That is, the contribution of the 
sterile component is determined by  the product 
$\sin \theta_{atm} \cdot \sin\beta$ 
and therefore 
\begin{equation}
\eta^{sun}_s = \sin^2 \theta_{atm} \sin^2 \beta
\label{etasunb}
\end{equation}
(instead of $\cos^2 \alpha$ in the (2 + 2)-scheme). 

Due to $\nu_{\tau} - \nu_s$ mixing 
the atmospheric $\nu_{\mu}$ will oscillate with 
$\Delta m^2 = \Delta m^2_{atm}$ into the state 
$\nu'$ (see Eq. (\ref{primestau})). According to (\ref{primestau}), 
the admixture of the sterile component is $\sin \beta$, so that  
\begin{equation}
\eta_s^{atm} = \sin^2 \beta. 
\label{etaatmb}
\end{equation}

From (\ref{etasunb}) and  (\ref{etaatmb}) we find that 
the total effect of the sterile components in the 
solar and atmospheric neutrino oscillations  can be described by 
\begin{equation}
\eta_s = \eta_s^{sun} + \eta_s^{atm} = \sin^2 \beta \cdot   
(1 + \sin^2 \theta_{atm}).
\label{31sum} 
\end{equation}
The parameter $\eta_s$ can be larger than 1 since in the $(3 +1)$-scheme
we make  
the double counting of the $\nu_s$ contributions  
from the two lightest states (they contribute both to $\eta_s^{sun}$ 
and to $\eta_s^{atm}$).  
In contrast with the (2 + 2)-scheme,  now $\eta_s$ is not fixed and it can
change from 
$(1 + \sin^2 \theta_{atm})$ to practically zero when 
$\sin^2 \beta \rightarrow 0$.  

According to  Eqs. (\ref{etaatmb}) and (\ref{etasunb}) 
\begin{equation}
\eta^{sun}_s = \sin^2 \theta_{atm} \eta^{atm}_s, 
\label{etasunb1}
\end{equation}
so that the inequality $\eta^{sun}_s \leq \eta^{atm}_s$ is the  generic
property of the $(3 +1)$-scheme under consideration. Taking 
$\sin^2 \theta_{atm} < 0.67$ (which corresponds to the lower bound 
$\sin^2 2\theta_{atm} > 0.88$ from the Super-Kamiokande data) we 
get from (\ref{etasunb}) and (\ref{boundatm}): 
$\eta^{sun}_s < 0.45$ 
independently on  solution of the solar neutrino problem. 
This bound is stronger than the immediate bound from the 
solar neutrino data.

\section{LSND oscillation probability in the $(3+1)$-scheme}
\label{sec:03}

Let us consider  predictions for the 
$\bar{\nu}_{\mu}\rightarrow\bar{\nu}_{e}$ oscillations relevant for 
the LSND result. The
short range oscillation results are determined by one $\Delta m^2_{LSND}$
only (flavour block is ``frozen'') and the oscillation pattern is
reduced to the $2\nu$ neutrino oscillations. So, one can use immediately
the results of the analysis of the LSND data in terms of the 
$2\nu$-oscillations. The depth of the oscillations or the effective mixing
parameter is given in Eq.~(\ref{eq:1b}). As in the one level 
dominance scheme~\cite{babu}, it is determined by the admixtures of
$\nu_e$ and $\nu_{\mu}$ in the isolated state. This prediction does
not depend on mixing in the flavour block, in
particular, on the solution of the solar neutrino problem.

Mixing matrix elements which enter  the expression for 
$\sin^2 2 \theta_{e \mu }$  (\ref{eq:1b}) are restricted 
by short baseline 
experiments. In the range of $\Delta m^2$ relevant for the LSND result
the best bound on $U_{e4}^2$ is given by the Bugey 
reactor experiment~\cite{bugey}. The $\nu_e$-survival probability in
this experiment is determined by
\begin{equation}
\sin^2 2 \theta_{ee}= 
4\, U_{e 4}^2 (1-U_{e 4}^2) \approx 4\, U_{e 4}^2 .
\label{eq:1bw}
\end{equation} 
The best bound on $U_{\mu 4}^2$, for $0.3$~eV$^2 < \Delta m^2<8$~eV$^2$,  
follows from the  $\nu_{\mu}$-disappearance searches in the 
CDHS experiment~\cite{cdhs}.
The relevant mixing parameter equals 
\begin{equation}
\sin^2 2 \theta_{\mu\mu}= 
4\, U_{\mu 4}^2 (1-U_{\mu 4}^2) \approx 4\, U_{\mu 4}^2. 
\label{eq:2bw}
\end{equation} 
From Eqs. (\ref{eq:1b}), (\ref{eq:1bw}) and (\ref{eq:2bw}) 
we get unique relation between the depths of oscillations in the 
CDHS, Bugey and LSND experiments: 
\begin{equation}
\sin^2 2 \theta_{e \mu} \approx 
\frac{1}{4} \sin^2 2 \theta_{e e} \cdot \sin^2 2 \theta_{\mu \mu}.  
\label{eq:rel}
\end{equation}
This relation 
holds in the one level dominance scheme when only one 
$\Delta m^2$ contributes to oscillations in all three experiments.  
The relation is modified in more complicated 
situations,  e.g., in schemes with more than one sterile neutrino. 

\begin{figure}
\begin{center}
\parbox[r]{4.5in}
{\mbox{\qquad\epsfig{file=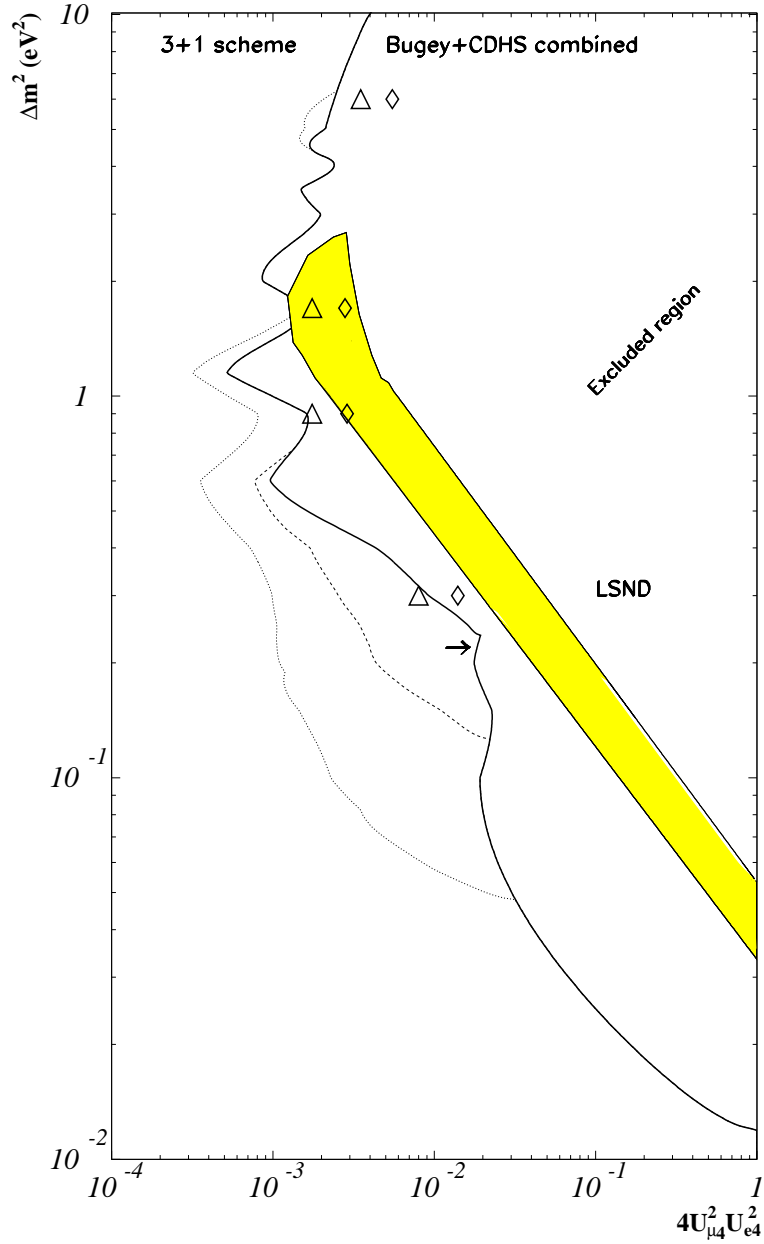,width=0.9\linewidth}}}
\end{center}
\caption{
The LSND allowed region (shadowed) and the 
bounds on the oscillation parameters in the (3 + 1) - scheme 
from the $\nu_{e} -$  and  $\nu_{\mu}- $ disappearance experiments. 
The lines show the limits obtained as the products of the 
90 \% C.L. upper bounds on $|U_{e4}|^2$ and $|U_{\mu 4}|^2$. Solid line
is  product of the bounds from  Bugey and CDHS  
(above the arrow) and is  product of the bounds from  Bugey and 
atmospheric neutrinos (below the arrow). Also shown are the bounds 
obtained as the product of bound from Bugey and expected bound from 
Mini-BooNE (dashed line) and BooNE (dotted line) experiments. 
The triangles and rombs show respectively the  95 and 99 \% C.L.  bounds
obtained in this paper (see Eq. {\protect \ref{eq:99c}}). 
}  
\label{fig10}
\end{figure}

The Bugey~\cite{bugey} and CDHS~\cite{cdhs} experiments have published
the $90\%$ C.L. upper bounds on $\sin^2 2 \theta_{ee}$ and 
$\sin^2 2 \theta_{\mu\mu}$ as functions of the $\Delta m^2$. These
bounds transfer immediately to the bounds on $U_{\mu 4}^2$ and  
$U_{e4}^2$:
\begin{eqnarray} 
U_{e   4}^2 & \lsim U^2_{Bugey}(\Delta m^2), \\ \nonumber
U_{\mu 4}^2 & \lsim U^2_{CDHS}(\Delta m^2). 
\label{eq:3bw}
\end{eqnarray} 
Using Eq.~(\ref{eq:1bw}) and Eq.~(\ref{eq:2bw}) one gets the 
upper bound on $\sin^2 2 \theta_{e \mu}$~\cite{giunti,barger}:
\begin{equation}
\sin^2 2 \theta_{e \mu }= 
4\, U_{\mu 4}^2 U_{e 4}^2 \lsim
4\, U^2_{Bugey}(\Delta m^2)\cdot U^2_{CDHS}(\Delta m^2) .
\label{eq:4bw}
\end{equation} 
The bound (\ref{eq:4bw}) shown in Fig. \ref{fig10} excludes most of the region
of parameters indicated by the LSND experiment. On this
basis, it was
concluded that $(3+1)$ -mass scheme  cannot reproduce the 
LSND result~\cite{giunti,barger}. 
The question is,  however, which  confidence level  should be ascribed
to the bound (\ref{eq:4bw})? Or,  at which  confidence level the LSND
result is excluded by  the combined Bugey and
CDHS bounds? 

Let us introduce the probabilities, $P_{\alpha}(U_{\alpha4}^2, \Delta m^2)$,
$\alpha=e,\mu$, that experimental data from Bugey or  CDHS 
correspond to a given value of 
$U_{\alpha4}^2$ for fixed $\Delta m^2$. The probabilities 
$P_{\alpha}(U_{\alpha4}^2, \Delta m^2)$ should be found from the fit of
experimental data. In such a fit  $U_{\alpha4}^2$  
can take both positive
and negative values but with $|U_{\alpha4}^2|<1$.  Therefore the  
normalization conditions  for  $P_{\alpha}$ should be written as  
$\int_{-\infty}^{\infty} dx P_{\alpha}(x, \Delta m^2)=1$. The  
$90\%$ C.L. bounds on $U_{\alpha4}^2$ are determined by the condition
\begin{equation}
\int_{-\infty}^{U^2_{\alpha 4}} dx
P_{\alpha}(x, \Delta m^2)=0.9~.
\label{eq:99a}
\end{equation}

The probability that the product of $U_{e4}^2 \cdot U_{\mu4}^2$ is
larger than certain value $\rho$ is given by the integral
\begin{equation} 
{\cal P}(\rho) \equiv 
\int_{\rho}^{\infty} dy
\int_{\rho/y}^{\infty} dx
P_{e}(x, \Delta m^2)
P_{\mu}(y, \Delta m^2).
\label{eq:99c}
\end{equation}
The $95\%$  and $99\%$ C.L. upper
bounds, $\rho_{95}$ and $\rho_{99}$, are determined by the conditions
\begin{equation}
{\cal P}(\rho=\rho_{95}) = 0.05, ~~~~~~{\cal P}(\rho=\rho_{99})=0.01~. 
\label{conflev}
\end{equation}

It is impossible to reconstruct 
the probabilities, $P_{\alpha}(U_{\alpha4}^2, \Delta m^2)$, 
from published $90\%$ C.L. exclusion plots. 
Therefore to make estimations,  we adopt the following procedure:  

\noindent
1). We assume
that the distributions  $P_{\alpha}(U_{\alpha4}^2, \Delta m^2)$ have
the  Gaussian form. The latter is  characterized by central value,
$\bar{U}_{\alpha4}^2$, and by the widths $\sigma_{\alpha}$. 

\noindent
2). For several  $\Delta m^2$  the
experimental groups have published the best fit values of
$U_{\alpha4}^2$ and we use them as the central values 
$\bar{U}_{\alpha4}^2$ in the Gaussian distributions.  

\noindent
3). For those $\Delta m^2$ we find the widths of the distributions using
Eqs.~(\ref{eq:99a}) and the published $90\% $ C.L. upper bounds on
$U^2_{\alpha 4} = 1/4 \sin^2 2\theta_{\alpha \alpha}$. 

\noindent
4). Using definitions (\ref{conflev}) and  Eq.~(\ref{eq:99c}),  we
calculate the 
$95\% $ C.L. and $99\%$  C.L. bounds,  
on the products of the matrix elements, and consequently, 
on  $\nu_e - \nu_{\mu}$ mixing parameter:  $\sin^2 2\theta_{e \mu} \leq
4\rho$.

The bounds on $\sin^2 2\theta_{e \mu}$ are shown in
Fig.~\ref{fig10} for different values of 
$\Delta m^2$ by rombs and triangles. Also shown,  is the $90\%$ C.L.
allowed region of LSND  taken from Ref.~\cite{eitel}. 
As follows from the figure, in the interval  
$\Delta m^2 = 0.4 - 2$~eV$^2$, some part
of the  LSND region is compatible with the Bugey 
and CDHS bounds  at $95 - 99\%$ C.L. . 
The upper limit given by the
product of the bounds  corresponds to $95\%$ C.L. in the range of
$\Delta m^2\lsim 1$ ~eV$^2$  and it approaches $99\%$ C.L. for
smaller values of $\Delta m^2$. 

For  $\Delta m^2 < 0.7$ eV$^2$, the bound from CDHS disappears and 
stronger restriction on $\sin^2 2\theta_{\mu \mu}$ 
follows from the    atmospheric 
neutrino results, namely,  from the up-down asymmetry which becomes 
suppressed when large admixture of $\nu_{\mu}$ exists in 
the heavy state~\cite{giunti}. 

Thus, the LSND result is 
consistent with bounds on mixing from other experiments at a few per
cent level in the  of the $(3 +1)$-scheme.

Introduction of more than 1 sterile neutrino can  enhance  the LSND
probability. The enhancement consistent with   
bounds on the $\nu_{e}$ -  and  $\nu_{\mu}$ - disappearance 
is however rather weak: about 20 \% (see Appendix).

\section{Improving the bounds on $\nu_{\mu}$- and $\nu_{e}$-  
disappearance}
\label{sec:02} 

Explanation of the LSND result in the $(3+1)$-scheme implies that both
$U_{e4}^2$ and  $U_{\mu4}^2$ are close to their present upper experimental
bounds. If $\nu_4$ consists mainly of the sterile component, the dominant
modes of oscillations will be 
$\nu_e \leftrightarrow \nu_s$ and $\nu_{\mu} \leftrightarrow  \nu_s$. 
If $\nu_4$ contains significant admixture of the  $\nu_{\tau}$, 
one should expect also $\nu_{\mu} \leftrightarrow \nu_{\tau}$ and 
$\nu_{e} \leftrightarrow \nu_{\tau}$ oscillations~\cite{giunti-31}. 
Therefore, further (even moderate) 
improvements of sensitivities of the oscillation searches with respect to 
the  Bugey and CDHS sensitivities should lead to 
discovery of oscillations  both in $\nu_{e}$- and in $\nu_{\mu}$- 
disappearance. Negative results of these searches 
will exclude the oscillation interpretation of 
the LSND result in the $(3 + 1)$-scheme.

Let us consider   possibilities to improve
bounds  on  $\nu_{\mu}$-  and $\nu_e$-
oscillation disappearance searches in the forthcoming experiments.

\subsection{$\nu_e$- disappearance experiments}

The $\bar{\nu}_e$- disappearance due to oscillations with 
$\Delta m^2 = \Delta m^2_{LSND} = (0.4 - 10)$ eV$^2$  can be 
searched for in the high statistics short base-line 
reactor experiments 
of the Bugey type. Keeping in mind restricted power of the reactors 
the only possibility to significantly increase statistics is to increase
the size  of the detector placed  close to the reactor. 
Notice that the absolute value of the neutrino flux 
is known with an accuracy about 3 - 5 \%, and therefore 
to improve substantially the Bugey bound one should 
search for the non-averaged oscillation effects.  
Notice that for $\Delta m^2 \sim 1$ ~eV$^2$ and $E = 3$ MeV 
the oscillation length equals $l_{\nu} \sim 7 - 8$ m which is on the
border of the averaging regime.  
 
No experiments of this type are planned now,  and it seems,  
the Bugey bounds on $U_{e4}^2$ will not be improved before Mini-BooNE
will report  its results~\cite{mini-boone}. 

The reactor  project can be considered if Mini-BooNE will confirm the 
LSND result.  Such an experiment will allow  to disentangle the 
(2 + 2) and    $(3 +1)$-schemes. Indeed, in the (2 + 2)-scheme 
one expects small $\nu_{e}$-disappearance effect which can be 
described by the mixing parameter of the order  the LSND 
mixing parameter: 
$\sin^2 2\theta_{e e}= 4(U^2_{e3}+U^2_{e4}) \sim 2 \sin^2 2\theta_{LSND}$. 
That is, the positive effect of oscillations  in the short-baseline 
reactor experiment will exclude the (2 + 2)-scheme. \\

\subsection{$\nu_{\mu}$ - disappearance and  
KEK - front detector  experiment} 

Let us consider a possibility to improve 
bound  on $U_{\mu4}^2$ using KEK front detectors~\cite{k2kproposal}. 
At KEK  the high intensity $\nu_{\mu}$-flux with energies 0.5 - 3 GeV 
(maximum at 1 GeV)  is formed in the decay pipe of the length 
200 m. Two front detectors are situated at the distance about 100 m 
from the end of the pipe. At the Fine Grained Detector (FGD) 
one expects to observe about 9200 $\mu$ - like events for
$1. \,\,10^{-20}$ protons on target (P.O.T.) , whereas 
at the 1 kton water Cherenkov 
detector the number of events is 30 times larger.

One can search for the $\nu_{\mu}$-disappearance  by 
measuring the energy distribution of the 
$\mu$ - like events. Let us evaluate the sensitivity of such a study. 
We calculate the ratios of numbers of events in the 
energy bins $i$:  
$E_{\mu}^i \div E_{\mu}^i + \Delta E_{\mu}$ 
with  ($N_{osc}^i$) and without ($N_{0}^i$) oscillations: 
\begin{equation}
R_i \equiv \frac{N^i_{osc}}{N^{i}_{0}}~.  
\label{eq:3a}
\end{equation}
The ratios can be written in the following way: 
\begin{equation}
R_i \approx \frac{1}{N} 
\int_{E^i_{\mu}}^{E^i_{\mu} + \Delta E}  dE_{\mu} 
\int_{0}^{\infty}dE_{\mu}' f(E_{\mu}, E_{\mu}') 
\int_{E_{\mu}'}^{\infty} dE_{\nu} 
F(E_{\nu}) \sigma (E_{\nu}, E_{\mu}') \bar{P} (E_{\nu})~, 
\label{eq:4a}
\end{equation}
where $f(E_{\mu}, E_{\mu}')$ is the (muon) 
energy resolution function, $F(E_{\nu})$  
is the neutrino flux at the front detector~\cite{oyama00}, 
$\sigma(E_{\nu},E^i_{\mu})$ is the total 
inclusive cross section
of the $\nu_{\mu} N \rightarrow \mu X$ reaction~\cite{lipari}. 
It is dominated by
quasi-elastic cross-section with sub-leading contributions from
other reactions in the energy range of the KEK experiment
($E_{\mu}^{max}\lsim 3$~GeV). In our calculations we use the quasi-elastic
cross section only. 
$N$ is the normalization factor which equals the
integral (\ref{eq:4a}) with $\bar{P} = 1$.
In Eq.~(\ref{eq:4a}), 
$\bar{P}(E_{\nu})$ is the neutrino survival
probability averaged over the production region: 
\begin{equation} 
\bar{P}(E_{\nu}) = 
\frac{1}{N_P(E_{\nu})}\int_{0}^{x_p} dx (L - x)^{-2} P(E_{\nu}, L - x)e^{-
\frac{\gamma
(E_{\nu})x}{\tau_{\pi}}}~.  
\label{eq:avprob}
\end{equation}
Here 
$$
\gamma (E_{\nu}) = \frac{E_{\nu}}{k m_{\pi}}
$$
is the effective Lorentz factor of pions which produce neutrinos with the 
energy $E_{\nu}$ and $k = 0.488 $, 
$m_{\pi}$ and $\tau_{\pi}$ are 
the mass and the lifetime of the pion; $L \approx 300$ m 
is the distance between the beginning of the decay pipe and the front   
detectors, $x_p$ is the length of the pipe. 
In Eq. (\ref{eq:avprob}) $N_P$ is the normalization factor which equals 
the same integral with ${P} = 1$. 

Since the energy resolution is rather good and the size of the 
energy bin is smaller than typical energy scale of 
the probability changes we omit the integration over 
$E_{\mu}$ and $E_{\mu}'$ and in the rest of integration take 
$E_{\mu}' = E_{\mu}^{i}$. 

The results of calculations of $R_i$ are shown in the 
Fig.~\ref{fig1b} for several values of  
$\Delta m^2$ and mixing angles at the border of the region 
excluded by the  CDHS experiment. 
The characteristic oscillation pattern is clearly seen 
for large $\Delta m^2$
$\gsim 6$ ~eV$^2$.  For  $\Delta m^2 \lsim 1$ ~eV$^2$, we find
very smooth distortion of spectrum with increasing deficit at events
of low energies. The deficit is at most $10\%$.  
As follows from the figure, with further increase of
statistics the  KEK front detector may  improve the CDHS bounds on 
$U_{\mu4}^2$  in the high $\Delta m^2$ part of the LSND allowed region. 

It would be 
interesting to estimate the possibility of water Cherenkov front
detector~\cite{oyama00} where the statistics should be  about 
$30$ times higher than in FGD. 
The problem here is that the muons 
with $E_{\mu} > 1.6$~GeV~\cite{chiaki}   are
not contained inside the detector~\cite{hill} and therefore 
the higher part of muon spectrum cannot be measured.

Let us underline that in this section we 
have described results of simplified estimations and 
the detailed study of possibilities by the experimental groups are
necessary.\\

\subsection{Mini-BooNE experiment,  and the neutrino mass
spectrum}

In the case of positive result of the 
${\nu}_{\mu} - {\nu}_{e}$ oscillation searches 
by the  Mini-BooNE we will have  
strong  evidence of existence of the sterile neutrino.  Moreover, the 
Mini-BooNE experiment itself may  allow us to disentangle the 
$(3 +1)$-  and (2 + 2)-schemes. 
Indeed, Mini-BooNE  can 
search for the $\nu_{\mu}$ - disappearance in the range 
$\Delta m^2 =  (8 \cdot 10^{-2} \div 20)$ ~eV$^2$. 
For $\Delta m^2 \sim 0.6 $ ~eV$^2$, the sensitivity to the 
effective mixing parameter can reach  0.15. 

In the (2 + 2)-scheme the $\nu_{\mu}$-disappearance driven by 
$\Delta m^2_{LSND}$  has  very small probability  
suppressed by  small admixtures of $\nu_{\mu}$ 
in the ``wrong" pair. If $\nu_{\mu}$ is mainly in the pair ($\nu_3$,
$\nu_4$) 
then  
\begin{equation}
\sin^2 2\theta_{\mu \mu} = 4 (U_{\mu 1}^2  + U_{\mu 2}^2)~.
\label{eq:boundmu}     
\end{equation}
It is at the level of mixing implied by the LSND: 
$\sin^2 2\theta_{\mu \mu} < \sin^2 2\theta_{LSND}$. 
 
In contrast, in $(3 +1)$-scheme
$\nu_{\mu}$-disappearance  should be at the level of  the bound
from the CDHS experiment~\cite{cdhs}, 
as is given in Eq. (\ref{eq:2bw}).  That is,  
$\sin^2 2\theta_{\mu \mu} \sim \sin^2 2\theta_{\mu e}/U_{e 4}^2$ 
-- the mixing parameter is enhanced by the factor $1 /U_{e 4}^2$.


If the Mini-BooNE experiment 
will find $\nu_{\mu}\rightarrow \nu_{e}$ oscillation effect  compatible
with LSND, but no $\nu_{\mu}$-disappearance will be detected, 
the $(3 +1)$-scheme will be excluded, provided that 
$\Delta m^2_{LSND} = (2 -  4) \cdot  10^{-1}$~eV$^2$. 

For $\Delta m^2_{LSND} > 4 \cdot 10^{-1}$~eV$^2$, the sensitivity of 
Mini-BooNE to $\nu_{\mu}$ disappearance is worse than the
present bound from CDHS experiment. Much better  sensitivity to
$\nu_{\mu}$ disappearance can be reached  
in the BooNE experiment. 
In Fig. \ref{fig10} we show the restrictions on the LSND 
oscillation mode from the bounds which can be
achieved by the Mini-BooNE and 
BooNE experiments instead of the CDHS bound.

\section{Phenomenology of $(3 +1)$ -mass scheme}
\label{sec:last}

In the ``minimal" version of the 
$(3 +1)$-scheme with normal  mass hierarchy and small
$\nu_s - \nu_{\tau}$  
mixing (one would expect $U_{\tau 4} \sim U_{\mu 4}$), the presence of the 
sterile neutrino in the flavor block is rather small: 
\begin{equation}
\sum_{i= 1,2,3} |U_{si}|^{2} \sim    U_{e 4}^2  +  U_{\mu 4}^2
\sim (3 - 10) \cdot 10^{-2}~.
\label{eq:a1}
\end{equation}
It will be difficult to detect such admixtures in oscillations 
driven by $\Delta m^2_{\odot}$ and $\Delta m^2_{atm}$, that is, 
in the solar,  atmospheric and the long base-line  experiments. 
Therefore, new elements of phenomenology of the scheme 
are associated mainly with the mixing of active neutrinos in the 4th
(isolated)  mass eigenstate.

\subsection{Neutrinoless Double beta decay}

\begin{figure}
\begin{center}
\parbox[c]{4.5in}
{\mbox{\qquad\epsfig{file=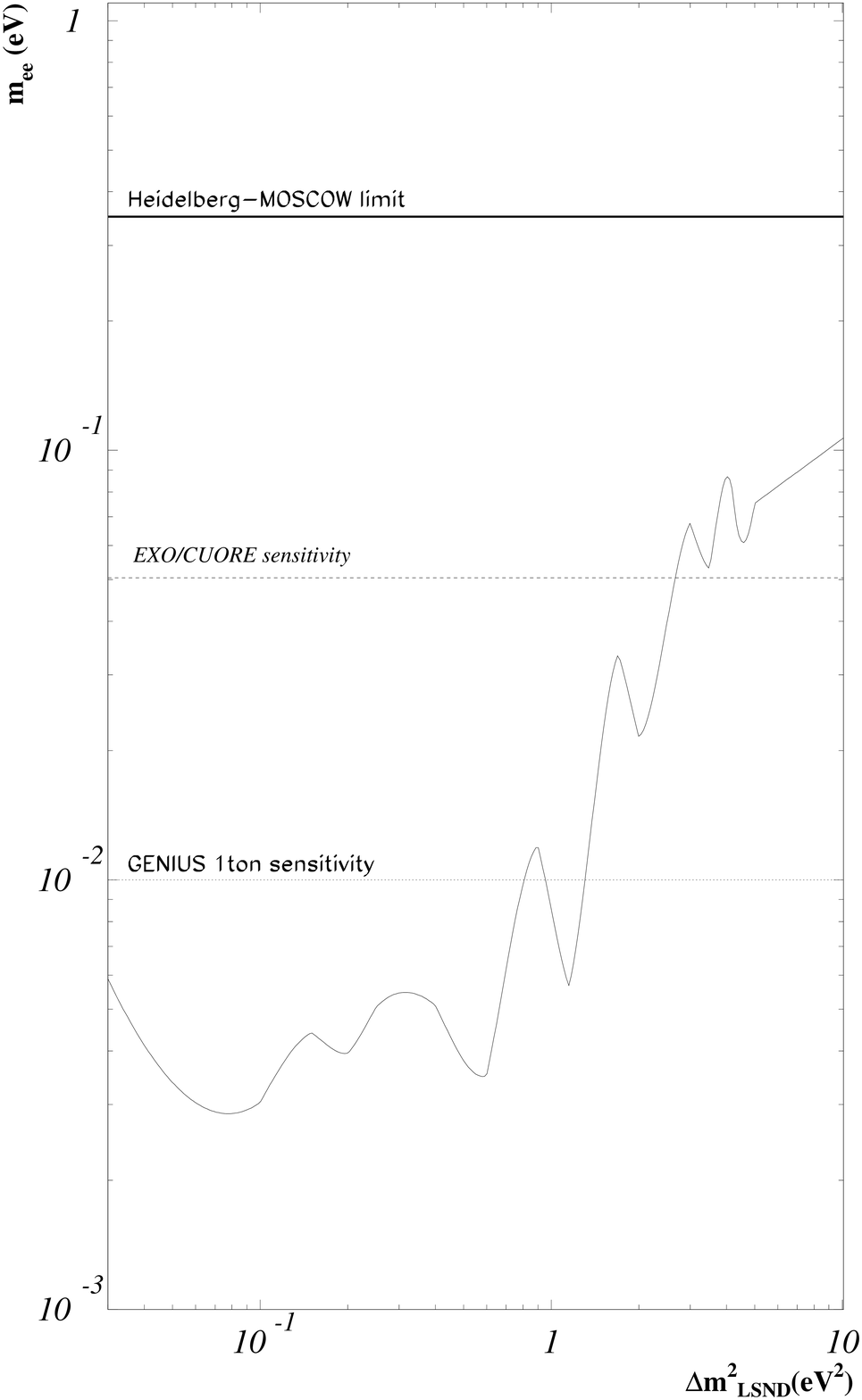,width=0.8\linewidth}}}
\end{center}
\caption{
The upper bound on the effective Majorana mass of the electron neutrino 
$m_{ee}$ as the function of $\Delta m^2_{LSND}$ in the $(3 +1)$ scheme 
with normal mass hierarchy. The bound corresponds to the upper limit on 
$|U_{e4}|^2$ from the Bugey experiment. Also shown are the present  
and future  limits on $m_{ee}$  from the 
$\beta \beta_{0\nu}$ -  experiments. 
}
\label{wfig1}
\end{figure}

\begin{figure}
\begin{center}
\hspace*{-2.0cm}
\parbox[c]{4.5in}
{\mbox{\qquad\epsfig{file=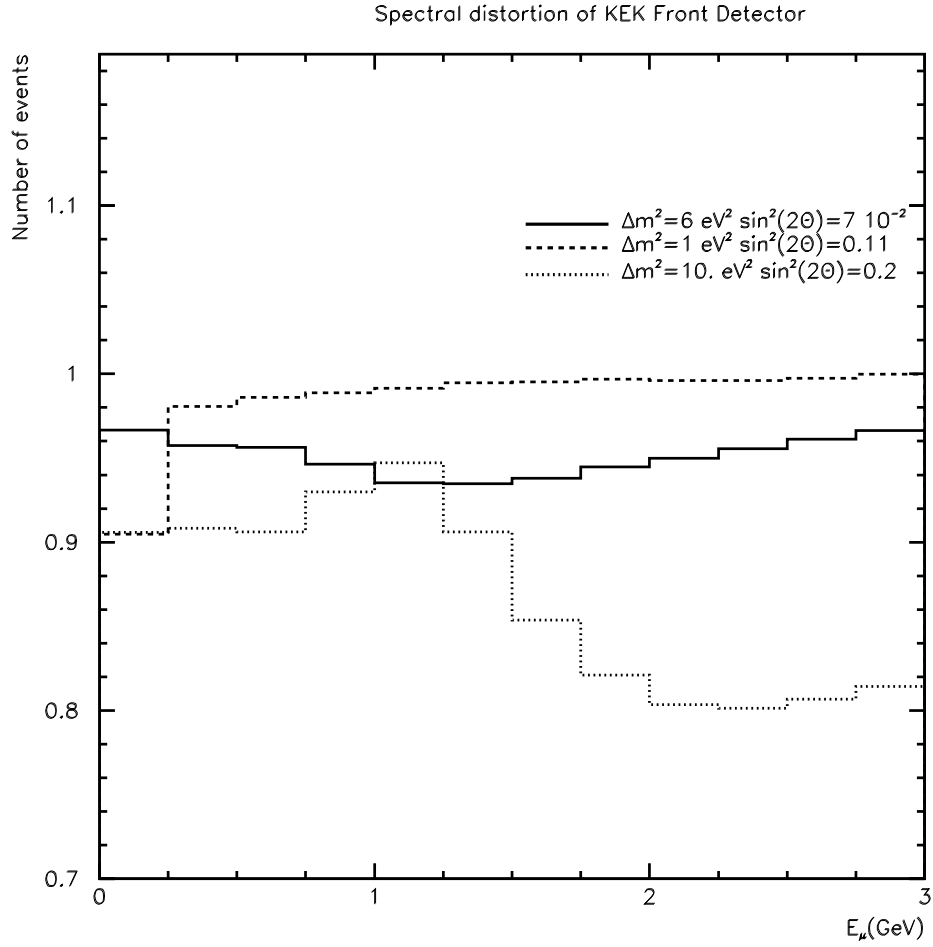,width=1.2\linewidth}}}
\end{center}
\caption{
Distortion of the energy spectrum of the $\mu$-like 
events  due to oscillations in the KEK - front detector experiment. 
Shown are the ratios 
of the predicted numbers of events with and without oscillations 
for   different values of the oscillation parameters. 
}  
\label{fig1b}
\end{figure}

If  neutrinos are the Majorana particles, the 
double beta decay, $\beta\beta_{0\nu}$, can put bounds on the mixing
parameters. Indeed, the (3 + 1)-scheme predicts 
the value for the effective Majorana mass of
the electron neutrino, $m_{ee}$, which can be accessible for the future
$\beta\beta_{0\nu}$  measurements. If the 
spectrum has normal mass hierarchy  with the isolated state being the
heaviest one, the dominant contribution follows from this heaviest 
state:
\begin{equation}
m_{ee} \sim m_{ee}^{(4)}\equiv U_{e4}^2 \sqrt{ \Delta m^2_{LSND}} 
\label{eq:last1}.
\end{equation}
The upper limit on $m_{ee}$ as  a function of $\Delta m^2_{LSND}$ in
the allowed region of the LSND result is shown in
Fig.~\ref{wfig1}. The Heidelberg-Moscow present limits on $m_{ee}$ 
is showed~\cite{betabeta}. As follows from the figure  the next generation
of $\beta\beta 0\nu$ experiments, the GENIUS (1 ton version)~\cite{genius}, 
the EXO and CUORE~\cite{exo} will be able to test this
scenario. 
For preferable values $\Delta m^2 \sim (1 - 2)$ ~eV$^2$ we get 
from (\ref{eq:last1}) $m_{ee} \sim (1 - 2) \cdot 10^{-2}$ eV. 

The $(3 + 1)$-scheme
with inverted mass hierarchy in which the three active
states form degenerate triplet with $m > (0.4 - 10)$~eV  
is restricted already by present data. 
The bound on $\beta\beta_{0\nu}$ rate~\cite{betabeta}  implies large
mixing angle solutions of the solar neutrino problem 
(LMA, LOW, VO) and cancellations of
contributions to $m_{ee}$ from different mass eigenstates~\cite{Dighe}.\\

\subsection{Atmospheric neutrinos}

The survival probability for the atmospheric muon neutrinos equals:
\begin{equation}
P_{\mu\mu}\equiv 1-2 U_{\mu4}^2 -4 U_{\mu3}^2 (1-U_{\mu3}^2-U_{\mu4}^2)
\sin^2 \Delta m^2_{atm},
\label{eq:1novo}
\end{equation}
where we have averaged over oscillations driven by the  mass split, 
$\Delta m^2_{LSND}$. According to Eq.~(\ref{eq:1novo}), the shape of the
zenith angle distribution and
the up-down asymmetry are determined by the effective mixing
parameter
\begin{equation}
\sin^2 (2 \theta^{eff}_{atm})= 
\frac{4\, U_{\mu3}^2 (1-U_{\mu3}^2-U_{\mu4}^2)}{1-2U_{\mu4}^2}\approx  
4\, U_{\mu3}^2 (1-U_{\mu3}^2+U_{\mu4}^2).
\label{eq:2novo}
\end{equation}
Notice that this parameter can be bigger than 1: e.g.,    
for $U_{\mu3}^2=1/2$ and $U_{\mu4}^2 = 2 \cdot  10^{-2}$ we get 
$\sin^2 2 \theta^{eff}_{atm}\approx 1.04$. This can
explain an  appearance of unphysical values of $\sin^2 2
\theta^{eff}_{atm}$ in the fits of the zenith angle
distributions. For small values of $\Delta m^2_{LSND}$, 
where the CDHS bound is weaker or absent, $U_{\mu4}$ can be large
enough and an enhancement of $\sin^2\theta_{atm}^{eff}$ 
can be significant. The presence of additional sterile state suppresses
the average survival probability
\begin{equation}
\langle P_{\mu\mu}\rangle 
\equiv 1-2 (U_{\mu4})^2 - 2U_{\mu3}^2(1-U_{\mu3}^2 - U_{\mu4}^2) .
\label{eq:3novo}
\end{equation}

\subsection{Nucleosynthesis bound}

Sterile neutrinos  are generated
in the Early Universe due to oscillations
$\nu_e\leftrightarrow \nu_s$ and $\nu_{\mu}\leftrightarrow \nu_s$ driven by
$\Delta m_{LSND}^2$ and mixing of $\nu_e$ and $\nu_{\mu}$
in the isolated 4th state. 
The mixing parameters for
$\nu_e\rightarrow\nu_s$ and  $\nu_{\mu}\rightarrow\nu_s$ channels,  
$\sin^2 2\theta_{es} \sim 4 U_{e4}^2$ and 
$\sin^2 2\theta_{\mu s} \sim 4 U_{\mu 4}^2$, are big
enough,  so that oscillations will lead to the 
equilibrium concentration of $\nu_s$. 
Thus in the epoch of primordial nucleosynthesis 
($T\sim$ few MeV)  
one expects the presence of four neutrino species in equilibrium. The
current limit,
$N_{\nu} < 3.4$,  can be avoided if large enough lepton asymmetry 
($\sim 10^{-5}$) existed already before oscillations driven by the
$\Delta m_{LSND}^2$  became efficient ($T\sim 10$ MeV). In this case,
the asymmetry  suppresses oscillations in the $\nu_e\rightarrow\nu_s$ and 
$\nu_{\mu}\rightarrow\nu_s$ channels. Notice however, that large
lepton asymmetry cannot be produced  in the scheme itself 
by the oscillations
to sterile neutrinos and anti-neutrinos: the mechanism~\cite{foot-volkas}
does not works. It requires the active neutrino state to be  heavier than
sterile state and very small mixing of sterile neutrinos in that  
heavier state.   
Another way to avoid the bound is to  assume that active neutrinos
were not in 
equilibrium before nucleosynthesis epoch~\cite{riotto}. In this case,
the number of additional  neutrino species can be even larger than 1.

\subsection{Supernova neutrinos}

Physics of the supernova neutrinos can be significantly modified by
presence of the sterile component. 

The difference of the  matter  potentials for
$\nu_e$ and  $\nu_s$ in electrically neutral  medium  equals
\begin{equation}
V_{es} = \sqrt{2} G_F n_e \left[1- \frac{n_n}{2n_e}\right] + V_{\nu}
=  \frac{G_F n}{\sqrt{2}} (3 Y_e - 1) + V_{\nu}~,
\label{poten}
\end{equation}
where $n$ is the total nucleon density; $Y_e \equiv n_e/n$
is the number of electrons per  nucleon;  $V_{\nu}$ 
is the potential due to neutrino-neutrino scattering which is important in
the region close to the neutrinosphere.   

\begin{figure}
\begin{center}
\parbox[c]{4.5in}
{\mbox{\qquad\epsfig{file=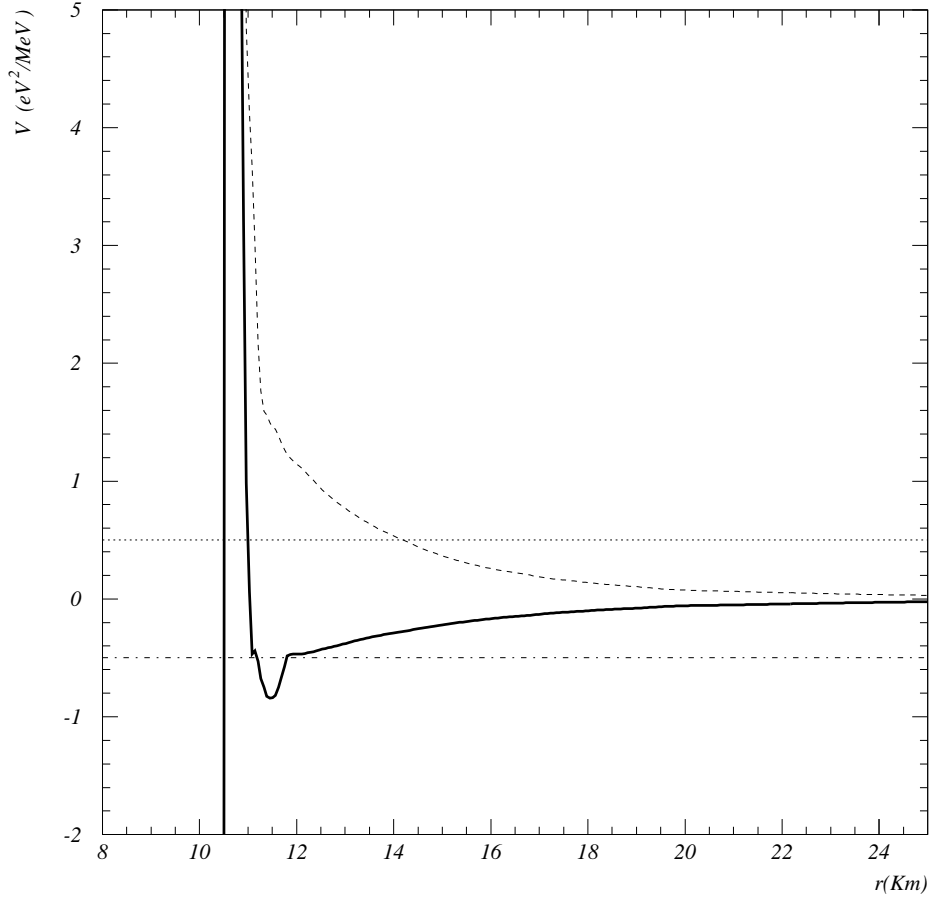,width=1.2\linewidth}}}
\end{center}
\caption{
The  matter potentials  for the 
$\nu_{e} - \nu_s$ system in the supernova as the function of distance from
the center of the star. Solid line shows the potential without
neutrino background 
and the dashed line -- with the background. The horizontal lines
correspond to $\pm \Delta m^2_{LSND}/2E_{\nu}$ for $E_{\nu} = 10$ MeV and     
$\Delta m^2 = 10$~eV$^2$; the positive (negative)  sign refers to 
neutrino (antineutrino) channel. Crossings of the potentials and the lines 
give the resonance points.
}
\label{wfig2}
\end{figure}

Typical dependence of the potential $V_{es}$ on
distance~\cite{3p1-sn,patel} is shown in Fig. \ref{wfig2} . For the
neutrino channel the  potential is negative
in the central parts; it becomes zero when $Y_e \approx  1/3$,  
then it changes the sign, reaches maximum and then falls down 
with density. The potential changes also with time: the gradients 
become larger. Moreover, after several seconds another 
region with  strong neutronization ($Y_e < 1/3$) 
may appear at densities  below $10^{7}$ g/cc.  
An  appearance of this region is related to the conversion 
$\nu_{e} \rightarrow \nu_s$  in the outer resonance 
(see later). As a consequence,  the reaction  
$\bar{\nu}_{e} + p \rightarrow e^+  + n$ will dominate  over 
${\nu}_{e} + n \rightarrow e^- +  p$,  thus producing the neutron reach 
region. It was suggested~\cite{fuller} that in this  region the
nucleosynthesis  of  heavy elements may take place. 

For the $\nu_{\mu} - \nu_s$ channel the difference of potentials 
equals
\begin{equation}
V_{\mu s} 
=  \frac{G_F n}{\sqrt{2}} (1 - Y_e) + V_{\nu}'. 
\label{poten2}
\end{equation}

Crossings of  the $\nu_{e}$- and $\nu_s$-  levels are  determined by the
condition: 
\begin{equation}
\frac{\Delta m^2_{LSND} \cos 2\theta}{2E} - V_{es} = 0~. 
\label{eq:res}  
\end{equation}
For  the anti-neutrino   channels 
$V_{es}$ or the first term should be taken with opposite  sign. 
Similar equation should be written for the potential 
$V_{\mu s}$. 

In Fig.~\ref{wfig2} the points of the resonances are shown as 
crossings of the $V_{e s}$ with the 
horizontal lines  given  by  
$\approx \Delta m^2/2E$ (lines below $V = 0$ correspond to 
the anti-neutrino channel).

From (\ref{poten}) and (\ref{eq:res}) we find 
the resonance densities (for $V_{\nu} = 0$): 
\begin{equation}
n = \frac{\Delta m^2_{LSND} \cos 2\theta }{2E} 
\frac{\sqrt{2}}{G_F (3Y_e - 1)}~. 
\label{eq:density}  
\end{equation}

\begin{figure}
\begin{center}
\hspace*{-4.0cm}
\parbox[c]{5.0in}
{\mbox{\qquad\epsfig{file= 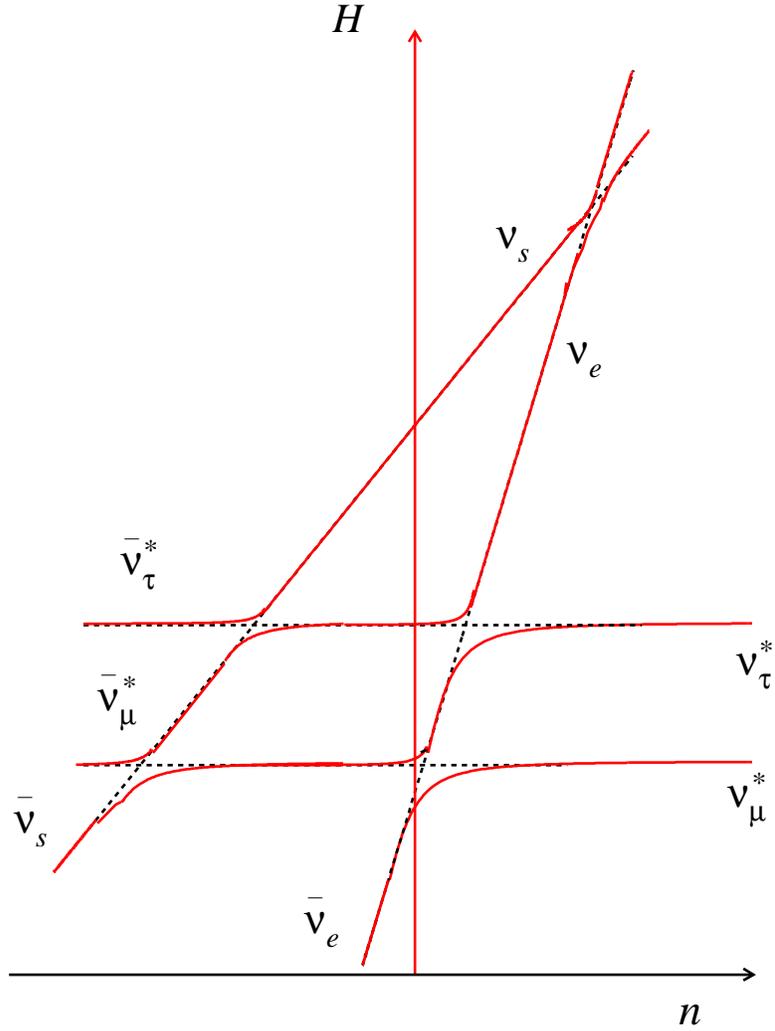,width=1.3\linewidth}}}
\end{center}
\vspace*{-3.9cm}
\caption{
The level crossing diagram  for the $(3 +1)$ scheme with normal 
mass hierarchy in supernova. Solid lines show the dependence of the 
eigenstates of the effective Hamiltonian on the matter density. 
Dashed lines show the dependence of the flavor states. The semi-plane 
with $n >0$ ($n < 0$) corresponds to the neutrino (antineutrino) 
channels. Only the outer resonance $\nu_e -\nu_s$ is shown.   
}
\label{xfig9}
\end{figure}

Using the potentials (\ref{poten}),(\ref{poten1}) 
and resonance conditions (\ref{eq:res})  we can construct  
the level crossing scheme.  
The scheme for the $(3 +1)$ spectrum with normal 
mass hierarchy is shown in fig.~\ref{xfig9}. 
We use the flavor basis $(\nu_s, \nu_e, \nu_{\mu}^*, \nu_{\tau}^*)$, 
where $\nu_{\mu}^*$, $\nu_{\tau}^*$ are the states which diagonalize 
the $\nu_{\mu}$ - $\nu_{\tau}$ sub-matrix of the total 
mass matrix. This rotation 
allows us to remove large mixing and it 
does not change  the physics since the produced $\nu_{\mu}$ and
$\nu_{\tau}$ fluxes are practically identical and  
they can not be distinguished at the detection.

The adiabaticity condition in the 
$\nu_{e}- \nu_s$ resonances  can be written as 
$\gamma >> 1$,  where the adiabaticity parameter equals   
\begin{equation}
\gamma = 4\pi^2 \frac{l_0 \cdot l_Y}{3 l_{\nu}^2} 
\sin^2 2\theta \left[Y_e + \frac{l_Y}{l_n}\left(Y_e - \frac{1}{3}\right)
\right]^{-1}~.  
\label{eq:adparam}
\end{equation}
Here $l_Y \equiv Y_e/(dY_e/dx)$,  $l_n \equiv n/(dn/dx)$, 
$l_{\nu} = 4\pi E/\Delta m^2$,  
$l_0 = \sqrt{2}m_N/G_F \rho$ are the vacuum oscillation length 
and  the refraction length.

In the case of normal mass hierarchy  
in the neutrino channels there are two  crossings of the 
$\nu_s$ and $\nu_e$ levels determined 
by the condition (\ref{eq:res}) (see Fig.~\ref{wfig2}). 
One (outer) crossing occurs at  relatively low density 
$\rho_r\sim 10^6 g/cm^3 (\Delta m_{LSND}^2/1$eV$^2$)~\cite{alei98}, 
when $Y_e$ substantially deviates from 1/3~ ($Y_e \sim 0.5$) (see
Fig.~\ref{xfig9}).  
It is characterized by the mixing parameter 
$\sin^2 2\theta_{es} \sim 4 U_{e4}^2$ 
($4 U_{e4}^2 \cos{\beta}^2$ in general). 
This resonance is highly adiabatic: the resonance density is
relatively small, 
so that the refraction length is  large. Also 
the gradients of the density and electron number are rather small. 
For $E \leq 40$ MeV, and 
$\sin^2 2\theta = 4 \cdot 10^{-2}$ 
we find that $\gamma \geq O(10)$ in the relevant range 
$\Delta m^2_{LSND}= (0.4 \div 10)$~eV$^2$. 

Thus, in the  outer resonance the  $\nu_{e}$   converts almost completely to   
$\nu_s$ with survival probability 
\begin{equation}
P_{ee} \approx U_{e4}^2 < (2 - 3) \cdot 10^{-2}~.  
\label{eq:survival}
\end{equation}
If $\Delta m_{LSND}^2 > 2 $~eV$^2$, the 
$\nu_e$-flux is absent above the layers with 
$\rho \sim (10^6-10^7) $g/cc, that is, in the region where $r$-processes
can lead to synthesis of heavy 
elements~\cite{fuller}. An absence of the $\nu_e$-flux enhances an
efficiency of the r-processes. For $\Delta m_{LSND}^2\lsim 1$ eV$^2$ the
effect of conversion on the $r$-processes is weak.

The second (inner) crossing occurs at much higher densities $\rho 
\sim 10^{11}$ g/cm$^3$~\cite{voloshin} in the layers with 
significant neutronization when  
$$
Y_e \approx 1/3
$$ 
(see Fig.~\ref{wfig2}). Here  the neutrino-neutrino scattering gives
substantial contribution 
to $V_{es}$ which  shifts the resonance. The level crossing occurs 
mainly due to 
change of the chemical composition (the ratio $n_n/n_e = n_n/n_p$). 
Taking $n \cdot m_N = 10^{10}$ g/cc, 
($m_N$ is the nucleon mass),  $l_Y = 3$ km, $E = 10$ MeV,  
$\Delta m^2 = 10$ ~eV$^2$ and  $\sin^2 2\theta_{e s} = 0.1$ we get 
for the adiabaticity parameter 
$\gamma \sim 0.3$, that is,  the
adiabaticity is broken.  For $\Delta m^2 > 30$ ~eV$^2$ the adiabaticity 
could be satisfied. This leads to disappearance of the 
$\bar{\nu}_{e}$-flux. Therefore observation of  the anti-neutrino
signal from SN1987A excludes such a possibility thus putting the upper
bound on $\Delta m^2$. 
(Notice that  the $\nu_{e}$-flux regenerates in 
the outer resonance  destroying conditions for the $r$-processes). 
At later time,  gradients of density and 
$Y_e$ become larger, in particular,     
$l_Y \sim 0.3$ km,  so that the adiabaticity breaks down even for 
large $\Delta m^2$.  

For $\Delta m^2 < 2$ ~eV$^2$ the adiabaticity is strongly broken for 
neutrinos of the detectable energies,  $E >  5$~MeV, during 
all the time of the burst and the effect of the inner
resonance can be neglected.

Let us consider properties of the neutrino bursts
arriving at the  Earth detectors.

1).  The generic prediction of the scheme is disappearance of the
``neutronization peak" due to $\nu_e \rightarrow \nu_s$
conversion in the outer resonance (\ref{eq:survival}) . 
The peak should be seen  neither 
in $\nu_{e}$ nor in $ \nu_{\mu}/ \nu_{\tau}$. 
Observation of the peak in the 
$\nu_{e}$ luminosity at the initial stage will exclude the scheme 
with normal mass hierarchy. Only if the inner resonance is adiabatic the
scheme can survive. 

2). During the cooling stage one expects hard $\nu_e$-spectrum 
which coincides with original spectrum of $\nu_{\mu}$ and/or
$\nu_{\tau}$:  $F(\nu_e) \propto  F^0(\nu_{\mu})$. 
The absolute value of the $\nu_e$-flux  depends on the mixing
in the flavour block. If $U_{e3}^2> 10^{-3}$, so that the level 
crossing related to 
$\Delta m^2_{atm}$ is adiabatic, one gets $F(\nu_e) =  F^0(\nu_{\mu})$
independently on solution of the solar neutrino problem. 
No Earth matter effect should be observed. 
If $U_{e3}^2 < 10^{-5}$, the level crossing is strongly non-adiabatic and
the $\nu_e$-signal depends on properties of the low density
resonance related to the $\Delta m^2_{\odot}$. 
For the LMA solution one gets: $F(\nu_e)=\cos^2 \theta_{\odot} 
F^0(\nu_{\mu})$. The flux can be larger in the case of SMA (small
mixing angle) solution depending on the level of adiabaticity.
Intermediate
situation is possible, if $U_{e3}^2$ is in the interval $10^{-5}-10^{-3}$.

In the antineutrino channels, the $\bar{\nu}_s$-level 
does not cross $\bar{\nu}_e$ level but it 
has  two crossings with $\bar{\nu}_{\mu}^*$
and $\bar{\nu}_{\tau}^*$ levels which can be both adiabatic. Still the
presence of the sterile neutrino will modify the $\bar{\nu}_e$
signal. The $\bar{\nu}_e$ propagation is adiabatic so 
that $\bar{\nu}_{e} \rightarrow \bar{\nu}_1$.  
If SMA solution is correct, one gets 
$\bar{\nu}_1 \approx \bar{\nu}_{e}$, that is $\bar{\nu}_{e}$-flux is
practically unchanged. For LMA (or LOW) the soft part of the spectrum
will be suppressed: 
\begin{equation}
F(\bar{\nu}_{e}) = \cos^2 \theta_{\odot} F^0(\bar{\nu}_{e})
\label{eq:soft}
\end{equation}
and the total $\bar{\nu}_{e}$ flux will depend on  characteristics  of
$\bar{\nu}_s - \bar{\nu}_{\mu}/\nu_{\tau}$ resonances. 
At least one of these resonances 
should be adiabatic since the mixing of $\bar{\nu}_s$ in the active block 
is $\sum_{i = 1}^3 U_{si}^2 \geq U_{e 4}^2 + U_{\mu 4}^2$ and the later 
is large  enough.  
If $\bar{\nu}_s$ crossing with $\bar{\nu}_{\mu}^*$ is adiabatic, then 
$\bar{\nu}_{\mu}^* \rightarrow \bar{\nu}_s$ and the $\bar{\nu}_{e}$ 
flux will be as in Eq. (\ref{eq:soft}). If the resonance is completely 
non-adiabatic, then $\bar{\nu}_{\mu}^* \rightarrow \bar{\nu}_2$, and since 
$\bar{\nu}_2$ has a component $\sin^2 \theta_{\odot}$ of the
$\bar{\nu}_{e}$-flux we get 
$F(\bar{\nu}_{e}) = \cos^2 \theta_{\odot} 
F^0(\bar{\nu}_{e})  + \sin^2 \theta_{\odot} F^0(\bar{\nu}_{\mu})$. 
In these cases also the Earth matter effect should be observed.

Some combination of the  $\bar \nu_{\mu}$ and $\bar \nu_{\tau}$
fluxes, will be converted to the sterile neutrino flux. 

In the case of inverted mass hierarchy, when the isolated fourth state is the
lightest one, the crossings of $\bar{\nu}_e$ and $\bar{\nu}_s$
levels 
appear in the antineutrino plane. As the result,
originally produced  $\bar \nu_{e}$ will be converted to $\bar \nu_{s}$. 
For $\Delta m_{LSND}^2 > 2$ ~eV$^2$ this will dump the efficiency of the
$r$-processes. 

The $\bar{\nu}_{e}$ flux at the Earth will be formed in the conversion
of $\bar \nu_{\mu}$ and $\bar \nu_{\tau}$ to $\bar \nu_{e}$ 
due to the mixing in the flavor block: more precisely, in the resonance 
associated with $\Delta m^2_{\odot}$. 
In the case of SMA solution this conversion is
practically absent, so that no $\bar \nu_{e}$-flux is expected at the Earth
in contradiction with observations of the $\bar \nu_{e}$-burst from
SN1987A. Therefore the  SN1987A data  require one of the large mixing
solutions. 
In the case of LMA or LOW solutions one expects 
 $F(\bar{\nu}_e)\sim \sin^2 \theta_{\odot} F^0(\bar{\nu}_{\mu})$. 
Properties of the $\nu_e$-flux are determined by mixing
in the flavour block~\cite{Dighe}.

\section{Discussions and conclusions}

1. Simultaneous explanation of the LSND result as well as the solar and
atmospheric neutrino data in terms of neutrino oscillations requires
introduction of additional (sterile) neutrino. In the favored 
(2 + 2) - scheme  the sterile component should be involved
significantly in the conversion of the solar and atmospheric neutrinos.

2. Recent experimental results
 give an indication that both in solar and in atmospheric
neutrinos the oscillation channels to active component dominate. We show
that in the (2 + 2)-scheme the equality 
$\eta_s^{sun} +  \eta_s^{atm} = 1$ 
should be fulfilled. The statistically significant deviation 
from this equality, 
in particular, a proof that  $\eta_s = \eta_s^{sun} +  \eta_s^{atm} < 1$
will lead 
to rejection of the (2 + 2)-scheme. 

In the case of  $(3 +1)$-scheme $\eta_s^{atm}$ can take any value  
between $\approx 0$ and $1 + \sin^2 \theta_{atm}$. 
The generic prediction of the scheme is that $\eta_s^{sun} \leq
\eta_s^{atm}$.

Determination  of $\eta_s$ from analysis of the solar and
atmospheric neutrino data  
will allow us to disentangle the (2 + 2) and $(3 +1)$-schemes.

We show that  the LSND result can be reconciled  with the bounds  on 
$\nu_{e}-$ (Bugey) and   $\nu_{\mu}-$ (CDHS) disappearance at 
95 \% - 99\% C.L.. The best agreement can be achieved at 
$\Delta m^2 = (1 - 2)$ ~eV$^2$.

4. The key consequences of the $(3 +1)$-scheme which explains the LSND
result  are that the  probabilities of 
the $\nu_{e} \leftrightarrow  \nu_s$  and 
$\nu_{\mu} \leftrightarrow  \nu_s$ oscillations with 
$\Delta m^2 = \Delta m^2_{LSND}$  should be at the level of present upper
bounds, so that  moderate improvements of the bounds should lead to discovery
of oscillations in these channels. We have estimated the possibility of
the KEK - front detectors experiments to improve the bound on $\nu_{\mu}$ 
disappearance. We show that for $\Delta m^2 > 6$ ~eV$^2$   
searches of the distortion of the energy 
spectrum of the $\mu$-like events may have better sensitivity to
oscillations than CDHS has.  The Mini-BooNE experiment can improve the CDHS 
bound for $\Delta m^2 < 0.6$ ~eV$^2$ and the BooNE experiment will have 
better sensitivity than CDHS  for $\Delta m^2 < 1$ ~eV$^2$. 

At the moment, no experiment is planned to improve the Bugey  bound 
on $\bar{\nu}_{e}$-disappearance in the range $\Delta m^2 = 
\Delta m^2_{LSND}$. The relevant experiment could be the one with  
large size detector situated close to the reactor. Such an experiment  
will deserve serious consideration if Mini-BooNE confirms the 
LSND result or/and some indications of the sterile neutrino 
involvement are seen in the solar and atmospheric neutrino
experiments.

5. In the $(3 +1)$-scheme the effective Majorana mass of 
the electron  neutrino 
can be about $2\cdot 10^{-2}$ eV  for the normal mass hierarchy and 
at the level of the upper experimental 
bound in the case of inverted mass hierarchy.

6. The scheme can reproduce  an ``unphysical" value of the effective
mixing
parameter: $\sin^2 2\theta^{eff} > 1$ in the two neutrino analysis of the
atmospheric neutrino data.

7. The scheme predicts that a sterile neutrino was in the thermal
equilibrium
in the epoch of primordial nucleosynthesis unless ``primordial'' 
lepton asymmetry suppressed the oscillations of active to sterile
neutrinos. 

8. Presence of the sterile neutrino in the $(3 +1)$-scheme can substantially 
influence the physics of supernova neutrinos.  
The $(3 +1)$-scheme with normal mass hierarchy and 
$\Delta m^2 > 2$ ~eV$^2$ can realize mechanism of disappearance of the 
$\nu_{e}$-flux from the region with densities 
$(10^6 - 10^8)$ g/cc, thus creating necessary conditions for the 
nucleosynthesis of the heavy elements in the $r$-processes 
(see however~\cite{patel}). 

The scheme predicts disappearance of the neutronization peak 
 and hard $\nu_{e}$- spectrum during the cooling stage. 
The presence of sterile neutrino can manifest itself also as suppression of
the $\bar{\nu}_{e}$ flux without change of the spectrum.\\

Let us outline  further developments.

1). Before Mini-BooNE result, the progress will  be related to further 
searches for the sterile neutrino components 
in the solar and atmospheric neutrinos.
The experimental groups are requested to get   
bounds on $\eta_s^{sun}$ and $\eta_s^{atm}$ parameters from the one
$\Delta m^2$ fit  of their results. 

Important bounds  will come from the SNO experiment: 
namely, from measurements of the CC~event rate and  comparison 
of this rate with the neutrino-electron 
scattering rate, from searches for the Earth regeneration effects and the 
distortion of the spectrum, from measurements of the NC-event rates and
the  double ratio $[NC]/[CC]$. 

As far as the atmospheric neutrino data are concerned, further
accumulation of data on the zenith angle distributions 
and neutral current interactions will  strengthen the bound 
on $\eta_s^{atm}$. Here the distributions of the upward going
muons and  the events in the NC -  enriched samples,    
as well as measurements of $\pi^0$ events rate are of special interest. 

Still it is difficult to 
expect that the bound $\eta_s < 1$ will be established at the high
confidence level.

2). The Mini-BooNE experiment will give the key result which  will
determine further developments.  In the case of negative result, still
searches for the sterile 
neutrinos will be continued  with some  other motivation. The positive
result will give  strong confirmation of 
existence of the sterile neutrino, and identification of 
correct four  (or more) states scheme will be the major issue. 

Notice that values of the $\sin^2 2\theta_{e \mu}$ at the upper side of
the LSND allowed region will favour the (2 + 2)-scheme. 
Also restrictions on $\Delta m^2$ (say excluding of 
$\Delta m^2 > 1$ ~eV$^2$) will have important implications.  

Mini-BooNE itself and BooNE will search for also 
$\nu_{\mu}$-oscillation disappearance. Detection of the disappearance will 
favor the $(3 +1)$-scheme. 

The ORLAND experiment~\cite{orland}  will have sensitivity  
for  $\nu_{\mu}\rightarrow \nu_{e}$ oscillations  which will cover  the
LSND region.

3). Later progress will be related to searches for the  $\nu_{\tau}-$ 
appearance  and $\nu_{\mu}-$ disappearance in the long baseline
experiments.   
OPERA~\cite{opera}  will allow one to improve bound on 
the sterile neutrinos. MINOS experiment~\cite{minos} 
will provide a decisive check of presence of sterile neutrinos 
by measurements of NC/CC  ratio. 

Special high statistics reactor experiment to search for the 
$\bar{\nu}_{e}$-disappearance can be  considered. 

Future neutrinoless double
beta decays searches can give an additional check of the model 
being especially sensitive to scheme with inverted mass hierarchy.

\section*{Appendix. LSND and additional sterile neutrinos}

Let us make several remarks on  possibility to  get larger 
predicted LSND effect  by introducing additional  sterile neutrino
$\nu_s'$. 
In general an  introduction of additional neutrino state can enhance
the LSND probability and/or influence (relax)  bounds from the laboratory and
atmospheric neutrino experiments on the relevant mixing parameters 
$U_{e4}$ and $U_{\mu 4}$.

1).  To influence the prediction for LSND oscillation probability, the 5th
mass eigenstate (being mainly sterile neutrino) should have non-zero mixing
with both $\nu_e$ and $\nu_{\mu}$. If $\nu_5$ has admixture
of $\nu_e$ (or $\nu_{\mu}$) only, it will open new channels for
$\nu_e$-($\nu_{\mu}$-) disappearance, thus enhancing the bound on
$U_{e4}^2$ (or $U_{\mu4}^2$) from Bugey (CDHS) experiment. At the same
time, it will be no additional contribution to the  LSND probability
in comparation with the $(3+1)$-scheme. As the result, one gets stronger bound
on  possible value of  $\sin^2 2\theta_{e \mu}$,
\begin{equation}
\sin^2 2 \theta_{e \mu }<
4\, U^2_{Bugey}(\Delta m^2) U^2_{CDHS}(\Delta m^2) .
\label{sec1}
\end{equation}

2). The sensitivities of the Bugey, LSND and CDHS results to 
$\Delta m^2$ have the following hierarchy: 
$\Delta m^2_{Bugey} <  \Delta m^2_{LSND} <  \Delta m^2_{CDHS}$. 
Then depending on value of $m_{5}^2$ one can get different situations: 

(i) $m_{5}^2 < \Delta m^2_{Bugey}$: neither bounds on $U_{\alpha 4}^2$
nor   the LSND probability are changed. 

(ii) $\Delta m^2_{Bugey} < m_{5}^2 <  \Delta m^2_{LSND}$: 
No additional contribution to the LSND  appears. 
At the same time, 
new channel will be open for the $\nu_{e}$-disappearance  which leads 
to stronger bound on $U_{e 4}^2$. As a consequence, the LSND 
probability will be suppressed. 

(iii)  $m_{5}^2 >   \Delta m^2_{LSND}$: for mixing angles relevant for the
LSND at this condition both bounds on $U_{\alpha 4}^2$ will be modified
and  the LSND will get an additional contribution from the second 
$\Delta m^2$. 

The mass $m_{5}$ should not be very  close to $m_4$.   
The splitting should be 
resolved by the LSND experiment, that is:   
~$\Delta m_{54}^2 >>10^{-2}$~eV$^2$.   
Following the analysis of Ref.~\cite{giunti}, we find that in the 
degenerate case, 
~$\Delta m_{54}^2 \ll 10^{-2}$~eV$^2$, 
the depths of the $\nu_{\mu}$- and $\nu_{e}$-oscillations equal 
\begin{equation}
\sin^2 2\theta_{\mu\mu}=4\, \eta_{\mu} (1-\eta_{\mu}),
\label{eq:last5}
\end{equation}  
\begin{equation}
\sin^2 2\theta_{ee}=4\, \eta_{e} (1-\eta_{e}),
\label{eq:last4}
\end{equation}  
where
$\eta_{\alpha} \equiv  \sum_{i=4,5} U_{\mu i}^2$, $\alpha=e,\mu$.
The depth  of the $\nu_{\mu} \rightarrow\nu_{e}$ channel driven by
$\Delta m^2_{LSND}$ is 
\begin{equation}
\sin^2 2 \theta_{LSND}= 4\sum_{i=4,5} U_{\mu i}^2 U_{e i}^2 \leq 
4 \sum_{i=4,5} U_{\mu i}^2 \sum_{i=4,5} U_{e i}^2 =\eta_e\eta_{\mu} 
= \frac{1}{4}\sin^2 2\theta_{ee}  \times \sin^2 2\theta_{\mu\mu}~, 
\label{eq:last6}
\end{equation}
where we have used the Schwartz inequality. Comparing 
result in (\ref{eq:last6}) with  the one in 
(\ref{eq:rel}) we see that 
the LSND probability in the scheme with additional neutrino 
is smaller than in  the $(3 + 1)$-scheme. 

As an example, we  have considered the neutrino mass spectrum with 
(i) $\Delta m^2_{4i} \approx (1 - 2)$~eV$^2$, where index $i=1,2,3$
enumerates
the eigenstates from the flavour block, (ii) $\Delta m^2_{5i} \geq
8$~eV$^2$ and (iii) non-zero admixtures, $U_{e5}$ and $U_{\mu 5}$,  of
the $\nu_e$ and $\nu_{\mu}$ in the 5th state. So,  both necessary
conditions discussed above are satisfied. Oscillations
associated with the 5th state are averaged in the Bugey and
LSND experiments and they are at least partly averaged in the CDHS
experiment. 

In the mass range, $ \Delta m^2_{5i}> 8$~eV$^2$, 
the CDHS bound is rather weak and stronger bound 
on the $U_{\mu 5}^2$ follows 
from the CCFR experiment~\cite{ccfr85}. This  CCFR bound  
on $U_{\mu 5}^2$ is still  a factor of two weaker than the CDHS bound on 
$U_{\mu 4}^2$ at $\Delta m^2 = (1 - 2)$ ~eV$^2$.   
Taking also  $U_{e5}^2 \sim U_{e4}^2
\sim  U_{Bugey}^2/2$ we have found   
that the probability of 
$\bar{\nu}_{\mu} \leftrightarrow \bar{\nu}_e$ 
oscillations can be enhanced 
by a factor $1.25$ at most with respect to 
result of the $(3 + 1)$-scheme  (see Eq.~(\ref{eq:1b})).

Notice also that introduction of the second sterile neutrino can further
aggravate a situation with primordial nucleosynthesis.

\section*{Acknowledgments}
We are grateful to Y. Oyama, C. Giunti, J. Hill and C. Yanagisawa for useful
comments.  This work was supported 
by the European Union TMR network ERBFMRXCT960090.\vglue -0.4cm

\section*{Note added}
After this work was accomplished the paper~\cite{minos-note} has
appeared in which the possibility to constraint the $\eta_s^{atm}$
parameter to be down to $0.3$ using the energy distribution of the 
CC and NC events at MINOS was discussed.

\bibliography{4nu_npb1}

\newpage 

\newpage

\pagestyle{empty}
\newpage

\pagestyle{headings}
\newpage

\newpage 

\newpage

\end{document}